\documentclass[12pt]{article}
  \usepackage{epsf}

  \def\({\c c}
  \def\|{\'\i}
  
  \def\nl {\par \noindent }

  \newcommand{\VEV}[1]{\left\langle{#1}\right\rangle}

\begin {document}

  \begin{flushright}
  {\small
  SLAC-PUB-9137\\
  FERMILAB-Pub-02/020-T \\
   February 2002\\
  }
  \end{flushright}
  \medskip

  \begin{center}
  {\large\bf A UNITARY AND RENORMALIZABLE THEORY OF THE STANDARD
  MODEL IN GHOST-FREE LIGHT-CONE GAUGE } \footnote{Research
  partially supported by the Department of Energy under contract
  DE-AC03-76SF00515 and DE-AC02-76CH03000 (PPS)}

  \vspace{0.9cm}

{\large Prem P. Srivastava}$\,^{a,b,c}$ \footnote{E-mail:\quad
prem@uerj.br or
prem@slac.stanford.edu}

\medskip 

and \vspace{0.2cm}

\medskip 

{\large Stanley J. Brodsky}$\,^{c}$ \footnote{E-mail:\quad
sjbth@slac.stanford.edu}\\

\vspace{0.8cm}

$^{a}\,${\em \it Instituto de F\'{\i}sica,
Universidade do Estado de Rio de Janeiro, RJ 20550 } \\
$^{b}\,${\em \it Theoretical Physics Department,
  Fermilab, Batavia, IL 60510}\\
$^{c}\,${\em \it Stanford Linear Accelerator Center \\
 Stanford University,
Stanford, California 94309}
\end{center}

  \begin{center}
  {\bf Abstract}
  \end{center}

Light-front (LF) quantization in light-cone (LC) gauge is used to
construct a unitary and simultaneously renormalizable theory of
the Standard Model.  The framework derived earlier for QCD is
extended to the Glashow, Weinberg, and Salam (GWS) model of
electroweak interaction theory.  The Lorentz condition is
automatically satisfied in LF-quantized QCD in the LC gauge for
the free massless gauge field. In the GWS model, with the
spontaneous symmetry breaking present, we find that the 't Hooft
condition accompanies the LC gauge condition corresponding to the
massive vector boson.  The two transverse polarization vectors for
the massive vector boson may be chosen to be the same as found in
QCD.  The non-transverse and linearly independent third
polarization vector is found to be parallel to the gauge
direction.  The corresponding sum over polarizations in the
Standard model, indicated by $K_{\mu\nu}(k),$ has several
simplifying properties similar to the polarization sum
$D_{\mu\nu}(k)$ in QCD.  The framework is ghost-free, and the
interaction Hamiltonian of electroweak theory can be expressed in
a form resembling that of covariant theory, except for few
additional instantaneous interactions which can be treated
systematically.  The LF formulation also provides a transparent
discussion of the Goldstone Boson (or Electroweak) Equivalence
Theorem, as the illustrations show.

  \vfill

  \vfill
  \newpage

  \section{Introduction}
  \label{intrO}

The quantization of relativistic field theory at fixed light-front
time $\tau = (t - z/c)/\sqrt 2$, which was proposed by Dirac
\cite{dir}, has found important applications
\cite{bro,bpp,ken,pre} in gauge field theory,  string theory
\cite{Metsaev:2000yu}, and M-theory \cite{Susskind:1997cw}, and it
has become a useful alternative  tool for the analysis of
nonperturbative problems in quantum
chromodynamics~\cite{Brodsky:2001bx}.  Light-front quantization
has been employed in the nonabelian bosonization \cite{wit} of the
field theory of $N$ free Majorana fermions. The (nonperturbative)
degenerate vacuum structures, the $\theta$-vacua in the Schwinger
model and  their absence in the Chiral Schwinger model, were shown
\cite{pre1c,pre1} to follow  transparently  in the {\it front
form} theory,  along with the natural emergence in the former case
of  their continuum normalization.  Also the requirement of the
microcausality \cite{weif} implies that the LF framework is more
appropriate for quantizing \cite{prec} the self-dual (chiral
boson) scalar field.

LF quantization is especially useful for quantum chromodynamics,
since it provides a rigorous extension of many-body quantum
mechanics to relativistic bound states:  the quark, and gluon
momenta and spin correlations of a hadron become encoded in the
form of universal process-independent, Lorentz-invariant
wavefunctions \cite{bro}.  The LF quantization of QCD in its
Hamiltonian form thus provides an alternative to lattice gauge
theory for the computation of nonperturbative quantities such as
the spectrum as well as the LF Fock state wavefunctions of
relativistic bound states \cite{bpp}.

We have recently presented a systematic
study~\cite{Srivastava:2000cf} of light-cone (LC) gauge
LF-quantized QCD theory following the Dirac
method~\cite{dir1,wei1} and constructed the Dyson-Wick S-matrix
expansion based on LF-time-ordered products.  In our
analysis~\cite{Srivastava:2000cf} one imposes the light-cone gauge
condition as a linear constraint using a Lagrange multiplier,
rather than a quadratic form.  We then find that the LF-quantized
free gauge theory simultaneously satisfies the covariant gauge
condition $\partial\cdot A=0$ as an operator condition as well as
the LC gauge condition.  The resulting Feynman rule for the gauge
field propagator in the LC gauge is doubly transverse
\begin{equation}
\VEV{0|\,T({A^{a}}_{\mu}(x){A^{b}}_{\nu}(0))\,|0} = {{i\delta^{ab}}
\over {(2\pi)^{4}}} \int d^{4}k \;e^{-ik\cdot x} \; \;
{D_{\mu\nu}(k)\over {k^{2}+i\epsilon}}
\end{equation}
where
\[
 D_{\mu\nu}(k)=-g_{\mu\nu} + \frac
{n_{\mu}k_{\nu}+n_{\nu}k_{\mu}}{(n\cdot k)} - \frac {k^{2}}
{(n\cdot k)^{2}} \, n_{\mu}n_{\nu}, \quad\quad n^\mu D_{\mu \nu} =
k^\mu D_{\mu \nu} =0, \]
and $n_{\mu}$ is the null four-vector,
gauge direction.  Thus only physical degrees of freedom propagate.
The remarkable properties of (the projector) $D_{\nu\mu}$ provide
much simplification in the computations of loop amplitudes.  In
the case of tree graphs, the term proportional to $n_{\mu}n_{\nu}$
cancels against the instantaneous gluon exchange term.  The
renormalization constants in the non-Abelian theory can be shown
to satisfy the identity $Z_1=Z_3$ at one-loop order, as expected
in a theory with only physical gauge degrees of freedom.  The QCD
$\beta$ function computed in the noncovariant LC
gauge~\cite{Srivastava:2000cf} agrees with the conventional theory
result~\cite{gross,polit}.  Dimensional regularization and the
Mandelstam-Leibbrandt
prescription~\cite{Mandelstam:1982cb,Leibbrandt:1987qv,Bassetto:1984dq}
for LC gauge were used to define the Feynman loop
integrations~\cite{Bassetto:1996ph}.  Ghosts only appear in
association with the Mandelstam-Liebbrandt prescription.  There
are no Faddeev-Popov or Gupta-Bleuler ghost terms.

In this paper we shall extend our LC gauge -- LF quantization
analysis to the Glashow, Weinberg and Salam (GWS) model of
electroweak Interactions based on the nonabelian gauge group
$SU(2)_{W}\times U(1)_{Y}$~\cite{gws}. It contains a nonabelian
Higgs sector which triggers spontaneous symmetry breaking (SSB). A
convenient way of implementing SSB and the (tree level) Higgs
mechanism in the {\it front form} theory is
known~\cite{pre4,pre5,pre6}.  One separates the quantum
fluctuation fields from the corresponding
 {\it dynamical bosonic condensate }
(or zero-longitudinal-momentum-mode)
 variables,
{\it before} applying the Dirac procedure in order to construct
the Hamiltonian formulation\footnote{See Appendix A and Ref.
\cite{bpp} for references to other alternative discussions on
SSB.}.  This procedure by itself should determine in a {\it front
form} theory if the condensate variable is  a c- or a q- number
(operator). In the description of SSB they are shown to be
background constants.  In  the Schwinger model, in contrast, it is
shown \cite{pre1c} to be an operator.  Its occurrence in the model
is crucial for showing, also in the LF framework,  the degenerate
vacuum structure ($\theta$-vacua), known in the conventional
theory since long time ago.

The tree level Lagrangian of the GWS model written in terms of the
set of (tree level) parameters, for example, $(e, m_{W}, m_{Z},
m_{h}, m_{u}, m_{d}) $ is constructed and quantized on the LF.
The model has the underlying initial gauge symmetry even after we
rewrite it such that it bestows quadratic mass terms to some of
the vector bosons. One is thus required to fix the gauge even when
quantizing the theory in its spontaneously broken symmetry phase.
For example, in the unitary gauge the Goldstone fields are gauged
away, leaving behind only physical degrees of freedom.  The
resulting massive gauge field then carries the Proca propagator
for which $D_{\mu\nu}(k)\to \left[-g_{\mu\nu} + {k^\mu k^\nu\over
M^2}\right]$ in (1).  Because of the growing momentum dependence
of the gauge propagator, the power counting renormalizability of
the theory becomes very difficult to verify in this gauge.  't
Hooft, however, demonstrated it by inventing the renormalizable
$R_{\xi}$ gauges~\cite{rxi,pesk} and employing
gauge-symmetry-preserving dimensional regularization.  This
framework, however, requires one to include in the theory
Faddeev-Popov ghost fields, even in abelian theory, where the
ghost fields couple to physical Higgs field as well. Several
additional parameters $\xi^{\gamma}, \xi^{Z}, \xi^{W}$ are
introduced in the theory.  Their renormalization must be taken
also into account and the physical S-matrix elements should be
shown not to depend on them.

In contrast, in the LC gauge LF-quantized theory framework for GWS
model, there are no ghosts to consider, neither in the abelian nor in
the nonabelian case.  The massive gauge field propagator has good
asymptotic behavior, the massive would-be Goldstone fields can be taken
as physical degrees of freedom, and the proof of renormalizability
becomes straightforward.  Together with the previous work on QCD we thus
obtain the simultaneous realization of unitary and renormalizable gauge,
in our framework, for the Standard Model theory of strong and
electro-weak interactions.

We start by considering in Section 2 the simpler case of the
abelian Higgs model.  The ingredients introduced here will be used
later in the quantization of the non-abelian GWS model.  The tree
level Higgs Lagrangian when re-written in terms of the chosen tree
level parameters $e, M,$ and $m_{h}$ still has the underlying
gauge symmetry.  We construct the LF Hamiltonian framework in the
LC gauge, $ \, A_{-}=0$,  following closely the procedure adopted
in our paper on QCD.  In the present case, where the gauge field
mass {$\, M \,$}is generated by the Higgs mechanism, we find that
the operator 't Hooft condition, $\,\partial\cdot A= M\, \eta\,$,
where $\eta$ is the would-be Goldstone field which also acquires
the same mass, accompanies simultaneously the LC gauge condition.
This is in contrast to the case of massless QCD where we have
correspondingly the Lorentz condition.

The polarization vectors of the gauge field, which are all
physical, are constructed, and their simplifying properties are
discussed in detail.  The interaction Hamiltonian which carries
also an instantaneous term (derived in Appendix B) in the
LF-quantized theory is constructed. The Fourier transform of the
free gauge field, the propagators of the massive vector boson, the
would-be Goldstone field, and the Higgs boson are derived.  The LF
quantization of the GWS model, which contains a nonabelian Higgs
sector, is considered in Section 3.

Appendix B discusses a systematic procedure for constructing the
instantaneous interaction terms required in the LF quantized field
theory.  It is illustrated by considering the Yukawa theory, abelian
Higgs model, and QCD.  In our LF framework $A_{+}$ and $\psi_{-}$ are
nondynamical and dependent field components.  While taking care of the
dependency, but without removing away these variables, we are able to
recast the ghost free interaction Hamiltonian in a form close to that of
covariant gauge theory.  Despite a few additional instantaneous terms,
it is straightforward to handle them in the Dyson-Wick expansion
constructed in the LF-quantized theory.  The nice properties of the
gauge propagator turn it into a practical computational framework.

The Goldstone Boson (or Electroweak) equivalence
theorem~\cite{cornwal} becomes transparent in our framework.  Its
content is illustrated by the computation of Higgs bosons and top
quark decays in Section 4.  The computation of muon decay shows
the relevance of the instantaneous interactions for recovering the
manifest Lorentz invariance in the physical gauge~\cite{schweda}
theory framework.

A new aspect of LF quantization, is that the third polarization of
the quantized massive vector field $A^\mu$ with four momentum
$k^\mu$ has the form $E^{(3)}_\mu = {n_\mu M / n \cdot k}$.  Since
$n^2 = 0$, this non-transverse polarization vector has zero norm.
However, when one includes the constrained interactions of the
Goldstone particle, the effective longitudinal polarization vector
of the vector particle is $E^{(3)}_{\rm eff \, \mu}= E^{(3)}_\mu -
{ k_\mu \,k \cdot E^{(3)} / k^2}$ which is identical to the usual
polarization vector of a massive vector with norm $E^{(3)}_{\rm
eff }\cdot E^{(3)}_{\rm
eff }= -1$.  Thus, unlike the conventional quantization
of the Standard Model, the Goldstone particle only provides part
of the physical longitudinal mode of the electroweak particles.

  \section{ The Quantization of the Abelian Higgs Model in LC Gauge}


The implementation of spontaneous symmetry breaking (SSB) and the
tree level Higgs mechanism on the LF have been understood for some
time.  A convenient description of SSB, which is useful for
constructing \cite{pre5} the tree-level Lagrangian in the Higgs
model, is reviewed in Appendix A. The relevant differences in the
LF quantized theory in the presence of SSB, when compared with the
conventional theory treatment, may already be seen in the abelian
Higgs model discussed below. The results obtained here will be
utilized later in the quantization of the GWS model which carries
in it a non-abelian Higgs sector (Section 3).

The abelian theory is described by \begin{equation} {\cal L}=
-\frac{1}{4}\, F_{\mu\nu}F^{\mu\nu} + {\vert {\cal
D}_{\mu}\phi\vert}^{2} -V(\phi^{\dag} \phi) \end {equation}
where $\phi$ is a complex scalar field, ${\cal
D}_{\mu}= (\partial_{\mu} +ieA_{\mu}),$ and $V(\phi)= \mu^{2} \,
\phi^{\dag}\phi + \lambda (\phi^{\dag}\phi)^{2}$ with $\lambda >0 $ and
$ \mu^{2} < 0$.  We choose the bosonic condensate, $\,<0| \phi |0>=
v/{\sqrt 2}\,$, to be real and separate it from the fluctuation field
$\varphi$ \begin{equation} \phi(x)= \frac{1}{\sqrt 2}\,v + \varphi=
\frac{1}{\sqrt 2}\,\left([\,v+ \,h(x)\,] + i\eta(x)\,\right)
\end{equation} such that the real fields $h(x)$ and $\eta(x)$ carry
vanishing vacuum expectation values.

The tree level Lagrangian, when the SSB is present, may be rewritten as
\begin{eqnarray} {\cal L}&=& -\frac{1}{4}\, F_{\mu\nu}F^{\mu\nu}
+\frac{1}{2} M^{2} A_{\mu}A^{\mu}+\frac{1}{2}{(\partial_{\mu} \eta)^{2}}
+ M\, A^{\mu}\partial_{\mu}\eta + \frac{1}{2}{(\partial_{\mu} h)^{2}} -
\frac{1}{2} \, m^{2}_{h} \, h^{2} \nonumber \\ &+& e \, (h
\partial_{\mu}\eta-\eta \partial_{\mu} h)\, A^{\mu} + e\,M\,
\,A_{\mu}A^{\mu}\,h +\frac {e^{2}}{2} (
h^{2}+\eta^{2})\,A_{\mu}A^{\mu}\,\nonumber \\ &-& \frac{e\,
m_{h}^{2}}{2\,M}\,(\eta^{2}+ h^{2})\, h - \frac{\lambda}{4}\, (\eta^{2}+
h^{2})^{2}+ const.  \end{eqnarray} where $e, M, $ and $m_{h}$ indicate
the tree level parameters defined by $M =e\,v$, $m^{2}_{h}= 2\lambda
v^{2}=-2\mu^{2}$ indicating the physical squared mass of the Higgs field
$h(x)$, $2\,\lambda\, v=e\, m_{h}^{2}/M$, and $2\,\lambda= e^{2}\,
m_{h}^{2}/M^{2}$.  We note the presence of the mixed bi-linear term
involving the Goldstone field $\eta$ and the gauge field.

In view of the underlying local $U(1)$ gauge symmetry, one possible
choice of the gauge may be taken to be such that the Goldstone
mode $\eta$ is eliminated, the so called ``unitary (or unitarity)
gauge", where only the physical fields appear in the Lagrangian.
The gauge field is massive and its ( Proca) propagator falls off
more slowly than $1/k^{2}$ for large $k$.  The perturbation theory
renormalizability in this gauge is then not simple to demonstrate.
The alternative of ``renormalizability" or $R_{\xi}$ gauges were
introduced by 't Hooft~\cite{rxi}.  The gauge-fixing term is here
assumed to be ${\cal L}_{GF}= -(\partial\cdot A -\xi M
\eta)^{2}/(2\xi)$.  The bi-linear mixing of $\eta$ and $A_{\mu}$ is
then eliminated, and for any finite value of $\xi,$ all of the
propagators in this class of gauges fall off as $1/k^{2}$.  The
theory may also be shown~\cite{thooft} to be perturbatively
renormalizable.  We note, however, that in the Faddeev-Popov
quantization procedure we are required to introduce also the
auxiliary ghost fields in the theory with the corresponding piece
in the Lagrangian ${\cal L}_{\rm Ghost}= {\bar c} [-\partial\cdot\partial-
\xi M^{2}(1+h/v)] c $, which contains the coupling of ghost fields
with the physical Higgs field.  In the nonabelian theory there
are, in addition, the coupling of ghosts with the gauge field
resulting from the term ${\bar c}^{a} (-\partial\cdot {\cal
D})_{ab} c^{b} $.

In what follows we will quantize the {\it front form} theory
described by the Lagrangian (3) in the LC gauge where the ghost
fields are seen to decouple in both the nonabelian and abelian
theories.  The LF coordinates are defined as
$x^{\mu}=(x^{+}=x_{-}=(x^{0}+x^{3})/{\sqrt 2},\, x^{-}=x_{+}=
(x^{0}-x^{3})/{\sqrt 2},\, x^{\perp})$, where $x^{\perp}=
(x^{1},x^{2}) =(-x_{1},-x_{2})$ are the transverse coordinates and
$\,\mu=-,+,1,2$.  The coordinate $x^{+}\equiv \tau$ will be taken
as the LF time, while $x^{-}$ is regarded as the longitudinal
spatial coordinate.  The LF components of any tensor, for example,
the gauge field, are similarly defined, and the metric tensor
$g_{{\mu}{\nu}}$ may be read from $A^{\mu} B_{\mu}=
A^{+}B^{-}+A^{-}B^{+} -A^{\perp}B^{\perp}$.  Also $k^{+} $
indicates the longitudinal momentum, while $k^{-}$ is the
corresponding LF energy.
Note that the LF Minkowski space
coordinates are not related to the conventional ones, $(x^{0},
x^{1},x^{2},x^{3})$, by a finite Lorentz transformation.

We follow the arguments given in Ref.  \cite{Srivastava:2000cf}
and introduce auxiliary Lagrange multiplier field $B(x)$ carrying
the canonical dimension three.  The {\it linear} gauge-fixing term
$(B A_{-})$ along with the ghost term ${\bar c} (-\partial\cdot
{\cal D}_{-}) c $ are added to the Lagrangian (4) such as to
ensure the Becchi-Rouet-Stora~\cite{stora} symmetry of the action.
The relevant free field propagators are thus determined from the
following bi-linear terms in the action \begin{eqnarray} \int
d^{2}x^{\perp}dx^{-} && \biggl\{\,{1\over 2}\left[(F_{+-})^{2}-
(F_{12})^2 +2F_{+\perp} F_{-\perp} \right]+ B A_{-}\nonumber \\ &&
+ \frac{1}{2} M^{2} (2A_{+}A_{-}-A_{\perp}A_{\perp}) +
M(A_{+}\partial_{-} \eta + A_{-}
\partial_{+}\eta -A_{\perp}\partial_{\perp}\eta) \nonumber \\ &&
+(\partial_{+}\eta)(\partial_{-}\eta)- \frac{1}{2} \partial_{\perp}\eta
\partial_{\perp}\eta \nonumber \\ && +(\partial_{+}h)(\partial_{-}h)-
\frac{1}{2} \partial_{\perp} h \partial_{\perp} h -\frac{1}{2} m_{h}^{2}
h^{2} + \cdots \biggr\} \end{eqnarray} where we note that the fields
${A}_{+}$ as well as $B$ have no kinetic terms, and they enter in the
action as auxiliary Lagrange multiplier fields.

The canonical momenta following from (5) are $\pi^{+}=0$, $\pi_{B}=0$,
$\pi^{\perp}=F_{-\perp}$, $\pi^{-}=F_{+-}= (\partial_{+}
A_{-}-\partial_{-} A_{+})\,$, $\pi_{\eta}= (\partial_{-}\eta+ M A_{-})$,
and $\pi_{h}=\partial_{-}h$, which indicate that we are dealing with a
constrained dynamical system.  The Dirac procedure~\cite{dir} will be
followed in order to construct a self-consistent Hamiltonian theory
framework, which is useful for the canonical quantization and in the
study of the relativistic invariance.  The canonical Hamiltonian density
is \begin{eqnarray} {\cal H}_{c}&=&{1\over 2} ({\pi}^{-})^{2}+{1\over
2}(F_{12})^{2}-
A_{+}(\partial_{-}\pi^{-}+\partial_{\perp}\pi^{\perp}+M^{2} A_{-} +M
\partial_{-}\eta) + \frac{1}{2} M^{2} A_{\perp}A_{\perp}\nonumber \\ &&
+ M A_{\perp} \partial_{\perp}\eta +\frac{1}{2} \partial_{\perp} h
\partial_{\perp} h +\frac{1}{2} m_{h}^{2} h^{2} +\frac{1}{2}
\partial_{\perp}\eta \partial_{\perp}\eta - B A_{-} + \cdots
\end{eqnarray} The primary constraints are $\pi^{+}\approx 0$,
$\,\pi_{B}\approx 0$ and
$\,{\chi^{\perp}}\equiv\pi^{\perp}-\partial_{-}A_{\perp}+
\partial_{\perp}A_{-}\approx 0$, $\chi_{\eta}\equiv \pi_{\eta} -
\partial_{-}\eta -M A_{-} \approx 0$, and $\chi_{h}\equiv
\pi_{h}-\partial_{-}h \approx 0,$ where $\,\approx\,$ stands for the
{\it weak }equality relation.  We now require the persistency in $\tau$
of these constraints employing the preliminary Hamiltonian, which is
obtained by adding to the canonical Hamiltonian the primary constraints
multiplied by undetermined Lagrange multiplier fields.  In order to
obtain the Hamilton's equations of motion, we assume initially the
standard Poisson brackets for all the dynamical variables present in the
theory.

We are then led to the following secondary constraints \begin{eqnarray}
\Phi\equiv \partial_{-}\pi^{-}+\partial_{\perp}\pi^{\perp}+
M\partial_{-}\eta & \approx & 0, \nonumber \\ A_{-} & \approx & 0
\end{eqnarray} which are already present in (5) multiplied by Lagrange
multiplier fields.  Requiring also the persistency of $\Phi$ and $A_{-}$
leads to another secondary constraint \begin{equation}
\Psi\equiv\pi^{-}+\partial_{-}A_{+}\approx 0. \end{equation} The
procedure stops at this stage, and no more constraints are seen to arise,
since further repetition leads to equations which would merely determine
the multiplier fields.

We analyze now the nature of the LF phase space constraints derived
above.  In spite of the introduction of the gauge-fixing term, there
still survives a first class constraint $\pi_{B}\approx 0$, while the
other ones are second class.  An inspection of the equations of motion
shows that we may add~\cite{dir1} to the set found above an additional
external constraint $B\approx 0$.  This would make the whole set of
constraints in the theory second class.  Dirac brackets satisfy the
property such that we can set the constraints as {\it strong} equality
relations inside them.  The equal-$\tau$ Dirac bracket
$\{f(x),g(y)\}_{D}$ which carries this property is straightforward to
construct~\cite{dir1,wei1}.  Hamilton's equations now employ the Dirac
brackets rather than the Poisson ones.  The phase space constraints on
the light front:  $\pi^{+}= 0$, $A_{-}=0$, $\chi^{\perp}= 0$,
$\chi_{\eta}= 0$, $\chi^{h}= 0$, $\Phi=0$, $\Psi= 0$, $\pi_{B}=0$, and
$B=0$ thus effectively eliminate $B$ and all the canonical momenta from
the theory.  The surviving dynamical variables in LC gauge are found to
be $h$, $\eta$ and $A_{\perp},$ while $A_{+}$ is a dependent variable
which satisfies
$\partial_{-}(\partial_{-}A_{+}-\partial_{\perp}A_{\perp}-M \eta)=0$.

The canonical quantization of the theory at equal-$\tau$ is performed
{\it via} the correspondence $i\{f(x),g(y)\}_{D} \to
\left[f(x),g(y)\right],$ where the latter indicates the commutator (or
ant-commutator) among the corresponding field operators.  The
equal-LF-time commutators of the transverse components of the gauge
field are found to be $$\left[A_{\perp}(\tau,x^{-},x^{\perp}),A_{\perp'}
(\tau,y^{-},y^{\perp})\right] =i\delta_{\perp\perp'} K(x,y)$$ where
$K(x,y)=-(1/4)\epsilon(x^{-}-y^{-})\delta^{2}(x^{\perp}-y^{\perp})$.
The commutators are nonlocal in the longitudinal coordinate, but there
is no violation~\cite{prec} of the microcausality principle on the LF.
At equal LF-time, $(x-y)^2 = -(x^{\perp}-y^{\perp})^{2} <0 $, is
nonvanishing for $x^{\perp}\neq y^{\perp},$ but
$\delta^{2}(x^{\perp}-y^{\perp})$ vanishes for such spacelike
separation.  The commutators of the transverse components of the gauge
fields are physical, having the same form as the commutators of scalar
fields in the {\it front form} theory.  We find also \begin{eqnarray}
\left[\eta(\tau,x^{-},x^{\perp}), \eta (\tau,y^{-},y^{\perp})\right]
&=&i K(x,y) \nonumber \\ \left[\eta(\tau,x^{-},x^{\perp}), A_{\perp}
(\tau,y^{-},y^{\perp})\right] &=& 0 \end{eqnarray} and some other
nonvanishing ones \begin{eqnarray}
\left[{\partial_{-}}A_{+}(\tau,x^{-},x^{\perp}), \eta
(\tau,y^{-},y^{\perp})\right] &=&i \,{M} K(x,y) \nonumber \\
\left[{\partial_{-}}A_{+}(\tau,x^{-},x^{\perp}), A_{\perp}
(\tau,y^{-},y^{\perp})\right] &=& i \,{\partial_{\perp}} K(x,y)
\nonumber \\ \left[h(\tau,x^{-},x^{\perp}),
h(\tau,y^{-},y^{\perp})\right] =i K(x,y)\ . \end{eqnarray}

The structure of the commutators found in the LC gauge quantized theory
on the LF indicates that in our framework the 't Hooft (gauge)
condition, $\,\, \partial\cdot A - M \eta =0$, is simultaneously
incorporated as an operator equation, along with the LC gauge condition
$A_{-}=0$.  This is in parallel to the result shown
\cite{Srivastava:2000cf} in the earlier work on (massless) QCD where the
Lorentz condition was found to be automatically incorporated.  It gave
rise there to the doubly transverse gauge field propagator which
simplified greatly the computations of loop corrections and allowed for
a transparent discussion of the renormalization theory and unitarity
relations in the physical LC gauge.

The {\it reduced} free LF Hamiltonian density in LC gauge, on making use
of the constraints above, is shown to be \begin{equation}\label{eq:h0}
{H_{0}}^{LF} = {1\over 2}
(\partial_{\perp}A_{\perp'})(\partial_{\perp}A_{\perp'}) + \frac{1}{2}
M^{2} A_{\perp}A_{\perp}+ {1\over 2}
(\partial_{\perp}\eta)(\partial_{\perp}\eta) + \frac{1}{2} M^{2}
\eta^{2} + {1\over 2} (\partial_{\perp}h)(\partial_{\perp}h) +
\frac{1}{2} m_{h}^{2} h^{2} \end{equation} where the bi-linear cross
terms are eliminated due to the presence of the 't Hooft condition in
the framework.

The Hamilton's equations are found to lead to $ (\partial\cdot
\partial + M^{2}) A_{\mu}=0$, $(\partial\cdot\partial + M^{2}) \eta=0$ and
$(\partial\cdot\partial+ m_{h}^{2}) h=0$.  Taking into
consideration the commutators among the field operators as derived
above,  we may write the momentum space expansions of the free (or
interaction representation) field operators.  Following the
procedure parallel to that employed in Ref.
\cite{Srivastava:2000cf} we may write
\begin{equation} A^{\mu}(x)={1\over {\sqrt {(2\pi)^{3}}}} \int
d^{2}k^{\perp}dk^{+}\, {\theta(k^{+})\over {\sqrt {2k^{+}}}}\,
\sum_{(\alpha)} {E_{(\alpha)}}^{\mu}(k) \left[a_{(\alpha)}(
k^{+},k^{\perp}) e^{-i{ k}\cdot{x}} +a^{\dag
}_{(\alpha)}(k^{+},k^{\perp}) e^{i{ k}\cdot{ x}} \right ]
\end{equation} and \begin{equation} \eta(x)={1\over {\sqrt
{(2\pi)^{3}}}} \int d^{2}k^{\perp}dk^{+}\, {\theta(k^{+})\over
{\sqrt {2k^{+}}}}\, \left[b(k^{+},k^{\perp}) e^{-i{ k}\cdot{x}}
+b^{\dag }(k^{+},k^{\perp}) e^{i{ k}\cdot{ x}} \right ]\ .
\end{equation} Here $k^{2}= M^{2}$, $(\perp)=(1),(2)$,
$(\alpha)=(\perp), (3)$, $ a_{(\alpha)}(k)=a^{(\alpha)}(k)$, $\,
a_{(3)}(k) =-i b(k)$, and the nonvanishing commutator $\,
[a_{(\alpha)}(k), {a^{\dag}}_{(\beta)}(l)]$
$=\delta_{\alpha\beta}$ $\delta^{2}(k_{\perp}-l_{\perp})$
$\delta(k^{+}-l^{+})$.

The three physical polarization vectors $E_{(\alpha)}^{\mu}(k)=
E^{(\alpha){\mu}}(k)$ of the massive gauge field (the mass arising
through Higgs mechanism), satisfying $ E^{(\alpha)}_{-}(k)=0$, are
constructed as follows.  The two which are transverse to $k_{\mu}$ may
be taken to be the same as defined in the earlier work on QCD, viz,
\begin{equation} E^{\mu}_{(\perp)}(k)= E^{(\perp){\mu}}(k)= -
\,D^{\mu}_{\perp} (k) \end{equation} with \begin{equation}
D_{\mu\nu}(k)= D_{\nu\mu}(k)= -g_{\mu\nu} + \frac
{n_{\mu}k_{\nu}+n_{\nu}k_{\mu}}{(n\cdot k)} - \frac {k^{2}} {(n\cdot
k)^{2}} \, n_{\mu}n_{\nu}, \end{equation} where the null four-vector
$n_{\mu}$ indicates the gauge direction, whose components have been
chosen conveniently to be $\, n_{\mu}={\delta_{\mu}}^{+}$, $\,
n^{\mu}={\delta^{\mu}}_{-}$.  We note that $E^{(\perp)}_{+}=\,
k^{\perp}/k^{+}$, $E^{(\perp)}_{\perp'}=
g_{\perp\perp'}=-\delta_{\perp\perp'}$.  They are also transverse to the
gauge direction $n_{\mu}$.  The doubly transverse property
\cite{Srivastava:2000cf} was very useful in the loop computations in
QCD.  We have \begin{eqnarray} \sum_{{(\perp)}=1,2} E^{(\perp)}_{\mu}(k)
E^{(\perp)}_{\nu}(k)= D_{\mu\nu}(k) , &&\qquad \quad g^{\mu\nu}
E^{(\perp)}_{\mu}(k) E^{(\perp')}_{\nu}(k)= g^{\perp \perp'} \\
k^{\mu}E^{(\perp)}_{\mu}(k)=0, \qquad && \qquad \quad n^{\mu}
E^{(\perp)}_{\mu} \equiv E^{(\perp)}_{-}=0 \end{eqnarray} such
that they are {\it spacelike} 4-vectors.  The linearly independent
non-transverse third polarization vector for the massive vector
boson, in our LC gauge framework, is a {\it null} 4-vector being
parallel to the gauge direction \begin{eqnarray} E^{(3)}_{\mu}(k)=
E_{(3)\mu}(k)=-\frac{M}{k^{+}}\,n_{\mu}, && \qquad q\cdot
E^{(3)}(k)=- M \frac {q^{+}}{k^{+}}, \nonumber \\ k\cdot
E^{(\alpha)}(k)=-M\,\delta_{(\alpha) (3)}, && \qquad
E^{(3)}(k)\cdot E^{(\alpha)}(q)=0 \end{eqnarray} such that
\begin{equation} E^{(3)}_{\rm eff\,\mu}(k)\equiv E^{(3)T}=
E^{(3)}_{\mu}(k)-(k\cdot E^{(3)}(k))\, (k_{\mu}/k^{2}) =
E^{(3)}_{\mu}(k)+ M (k_{\mu}/k^{2})
\end{equation} is {\it spacelike} and transverse to $k_{\mu}$ with
$E^{(3)}_{\rm eff}(k)\cdot E^{(3)}_{\rm eff}(k)= - M^{2}/k^{2}=
-1$.

The sum over the three physical polarizations is given by $K_{\mu\nu}$
\begin{eqnarray} K_{\mu\nu}(k)&=&\,\sum_{(\alpha)}
E^{(\alpha)}_{\mu}E^{(\alpha)}_{\nu} =\,D_{\mu\nu}(k)+
\frac{M^{2}}{(k^{+})^{2}}\, n_{\mu} n_{\nu} \nonumber \\ &=&-g_{\mu\nu}
+ \frac {n_{\mu}k_{\nu}+n_{\nu}k_{\mu}}{(n\cdot k)} - \frac
{(k^{2}-M^{2})} {(n\cdot k)^{2}} \, n_{\mu}n_{\nu}, \end{eqnarray} which
satisfies:  $\,\, k^{\mu}\,K_{\mu\nu}(k)= (M^{2}/k^{+})\, n_{\nu}$ and
$\,\, k^{\mu}\, k^{\nu}\,K_{\mu\nu}(k)= M^{2}$.  We recall also
\cite{Srivastava:2000cf} \begin{eqnarray} D_{\mu\lambda}(k)
{D^{\lambda}}_{\nu}(k)= D_{\mu\perp}(k) {D^{\perp}}_{\nu}(k)&=& -
D_{\mu\nu}(k), \nonumber \\ k^{\mu}D_{\mu\nu}(k)=0, \qquad \quad &&
n^{\mu}D_{\mu\nu}(k)\equiv D_{-\nu}(k)=0, \nonumber \\ D_{\lambda\mu}(q)
\,D^{\mu\nu}(k)\, D_{\nu\rho}(q') &=&
-D_{\lambda\mu}(q)D^{\mu}_{\rho}(q').  \end{eqnarray}

The expansion of the transverse components of the gauge field is then
rewritten as \begin{equation}\label{eq:  field} A_{\perp}(x)=-A^{\perp}=
-{1\over {\sqrt {(2\pi)^{3}}}} \int d^{2}k^{\perp}dk^{+}\,
{\theta(k^{+})\over {\sqrt {2k^{+}}}}\, \left[a_{(\perp)}( k) e^{-i{
k}\cdot{x}} +a_{(\perp)}^{\dag}(k) e^{i{ k}\cdot{ x}} \right ]
\end{equation} which, together with the independent (would be Goldstone
) field $\,\eta ,$ describe the massive gauge field.  It is convenient
to also define the dependent gauge field component, $A_{+}$, by using
the 't Hooft condition, $\,\partial\cdot A\vert_{A_{-}=0}= M\eta$
incorporated in our LC gauge framework.  We find \begin{equation}
A_{+}(x)= -{1\over {\sqrt {(2\pi)^{3}}}} \int d^{2}k^{\perp}dk^{+}\,
{\theta(k^{+})\over {\sqrt {2k^{+}}}}\, \left[a_{(+)}( k) e^{-i{
k}\cdot{x}} +a_{(+)}^{\dag}(k) e^{i{ k}\cdot{ x}} \right ]
\end{equation} if $a_{(+)}=a^{(+)}$ is defined such that
\begin{equation} k^{+} a_{(+)}(k)\,=\, \left [k_{\perp}\, a_{(\perp)}(k)
-i M \,b(k) \right] =\, \left [k_{\perp} a_{(\perp)}(k) + M a_{(3)}
\right].  \end{equation} while we set $a_{(-)}(k)= a^{(-)}(k)=0$ in view
of $A_{-}=0$.  The following nonvanishing commutator is straightforward
to derive \begin{equation} \left[ a_{(\mu)}(k),
\,a^{\dag}_{(\nu)}(l)\right]=\, K_{\mu\nu}(k)
\delta^{2}(k_{\perp}-l_{\perp}) \delta(k^{+}-l^{+}) \end{equation} where
$\mu,\nu= -,+,\perp $.  Following the standard procedure, the free
propagator of the massive gauge field $A_{\mu}$ is found to be
\begin{equation}
\VEV{0\vert T \left(A_{\mu}(x)A_{\nu}(y)\right)\vert 0}= \frac
{i}{(2\pi)^{4}}\int d^{4}k \frac {K_{\mu\nu}(k)}{
(k^{2}-M^{2}+i\epsilon)} \, e^{-i \, k\cdot (x-y)}.
\end{equation} It does not have the bad high energy behavior found
in the (Proca) propagator in the unitary gauge formulation, where
the would-be Nambu-Goldstone boson is gauged away.  For $M\to 0$
it reduces to the doubly transverse propagator found
\cite{Srivastava:2000cf} in connection with the LF quantized QCD
in the LC gauge.

The Higgs field $h(x)$ commutes with other field operators, and
its propagator is $i / (k^{2}-{m_{h}}^{2}+i\epsilon)$.  The
commutation relations in (8) imply that the field $\eta$ has an
off-diagonal nonvanishing propagator with the component $A_{+}$,
viz, $\VEV{0\vert T \left(\eta(x)A_{+}(y)\right)\vert 0} \neq 0$.
The $\, \eta \, \eta\, $ propagator is given by $ \, i /
(k^{2}-M^{2}+i\epsilon)$.  If we use the ML prescription to handle
the $1/k^{+}$ singularity along with the dimensional
regularization, the general power-counting analysis becomes
available \cite{Srivastava:2000cf}.  The propagators in the
framework have good asymptotic behavior; the divergences
encountered are no worse than in QED.  The proof of perturbative
renormalizability in the LC gauge in the {\it front form}
quantized theory presented here may be given straightforwardly
along the lines performed earlier in the conventional
\cite{thooft} equal-time theory.  In view of the simplifying
properties of $K_{\mu\nu}$ (and $D_{\mu\nu}$ ), the absence of
ghost fields, and the availability of the power counting rules,
when we employ the dimensional regularization along with ML
prescription, the effort required in our framework is comparable,
as in the case of the previous work on QCD, to that in the
conventional theory computations.

Some comments on the polarization vectors in LC gauge are in
order.  With the restriction $\, E^{(\alpha)}_{-}=0$ there are only
three linearly independent polarization vectors\footnote{It is
easily shown that $n_{\mu}$, $n^{*}_{\mu}$,
$E^{(\perp)}_{\mu}(k)$, where $n^{*}_{\mu}=\delta^{-}_{\mu}$ is
the null vector dual to $n_{\mu}=\delta^{+}_{\mu}$ constitute a
convenient basis for 4-vectors in the LF theory.} as discussed
above.  $E^{(\perp)}_{\mu}(k)$ are transverse with respect to both
$n_{\mu}$ and $k_{\mu}$ while the non-transverse
$E^{(3)}_{\mu}(k)$ is parallel to the gauge direction $n_{\mu}$,
being equal to the sum of a transverse piece (T) ($\equiv E_{\rm
eff}$ ) and a longitudinal one (L), when referred to the 4-vector
$k_{\mu}$:  \begin{eqnarray} && E^{(\alpha)
T}_{\mu}(k)=(g_{\mu\nu}-\frac{k_{\mu}k_{\nu}}{k^{2}})\,
E^{(\alpha)\nu}(k), \qquad\quad k\cdot E^{(\alpha) T}(k)=0 \nonumber \\
&&E^{(\alpha) L}_{\mu}(k)=\frac{k_{\mu}}{k^{2}}\, (k\cdot
E^{(\alpha)}(k))= -M\, \frac{k_{\mu}}{k^{2}}\, \delta_{(\alpha)(3)},
\nonumber \\ k \cdot E^{(\alpha) L}(k)&=&k\cdot E^{(\alpha) }(k)= -M
\delta_{(\alpha)(3)} , \qquad \quad E^{(\alpha) T}(k)\cdot E^{(\beta)
L}(k)=0 \end{eqnarray} such that $\,E^{(\perp) T}_{\mu}(k)=E^{(\perp)
}_{\mu}(k)$, $\,E^{(\perp) L}_{\mu}(k)= 0$, $\,E^{(3) L}_{\mu}(k)=
-M\,(k_{\mu}/k^{2})$, $\,E^{(3) T}_{\mu}(k)=
M\,(k_{\mu}/k^{2}-n_{\mu}/k^{+})$, and $\,E^{(3)\,L, T}_{-}(k) \neq 0$,
$\,E^{(3)\,L}(k)\cdot E^{(3)\,L}(k)= M^{2}/k^{2}= + 1$.

The following analogous decomposition of $K_{\mu\nu}$ is useful in
computations
\begin{equation}
K_{\mu\nu}(k)= K^T_{\mu\nu}(k)+K^L_{\mu\nu}(k)
\end{equation}
where\footnote{ $K^L_{\mu\nu}(k)\neq \sum_{(\alpha)}
E^{(\alpha)L}_\mu(k)\,E^{(\alpha) L}_\nu(k)$.}
\begin{eqnarray}
K^{L}_{\mu\nu}(k)&=& (\frac{M^{2}}{k^{2}})
\,d_{\mu\nu}(k) \nonumber \\ K^{T}_{\mu\nu}(k)&=&
K_{\mu\nu}(k)-K^{L}_{\mu\nu}(k)= D_{\mu\nu}(k)+M^{2}\left(\frac
{n_{\mu}n_{\nu}}{(n\cdot k)^{2}}-
\frac{d_{\mu\nu}(k)}{k^{2}}\right) \nonumber \\ &=&
(k^{2}-M^{2})\, \left[\frac{d_{\mu\nu}(k)}{k^{2}}-\frac{n_{\mu}
n_{\nu}} {k^{+ 2}}\right]
\end{eqnarray}
where
\begin{equation}
d_{\mu\nu}(k) = - g_{\mu\nu} + \frac
{n_{\mu}k_{\nu}+n_{\nu}k_{\mu}}{(n\cdot k)}, \qquad
k^{\mu}\,d_{\mu\nu}(k)= \frac{k^{2}}{k^{+}} n_{\nu}, \qquad
k^{\mu}k^{\nu}\,d_{\mu\nu}(k) =k^{2}\ .
\end{equation}
They are
symmetric and some interesting properties are $ \,K^{L}_{\mu
-}(k)=K^{T}_{\mu -}(k)= d_{\mu-}(k)= 0$, $\,k^{\mu}\,
K^{T}_{\mu\nu}(k)=0$, $k^{\nu}\, K^{T}_{\mu\nu}(k)=0$,
$\,k^{\mu}\, K_{\mu\nu}(k)=k^{\mu}\, K^{L}_{\mu\nu}(k)=
(M^{2}/k^{+}) n_{\nu}$, $\,k^{\mu}k^{\nu}\,
K_{\mu\nu}(k)=k^{\mu}k^{\nu}\, K^{L}_{\mu\nu}(k)= M^{2} $.  From
the properties of $D_{\mu\nu}(k)$ we easily derive
\begin{equation} K_{\mu\rho}(k)K^{\rho}_{\nu}(k)=
\,d_{\mu\rho}(k)d^{\rho}_{\nu}(k) = -D_{\mu\nu}(k) \end{equation}
and \begin{eqnarray} K^{L}_{\mu \rho}(k)\,K^{T\rho}_{\nu}(k)&=& -
\,\frac {M^{2}(k^{2}-M^{2})} {(k^{2})^{2}}\, D_{\mu\nu}(k),
\nonumber \\ K^{L}_{\mu \rho}(k)\,K^{L\rho}_{\nu}(k)&=& - \,\frac
{M^{4}} {(k^{2})^{2}}\, D_{\mu\nu}(k), \nonumber \\ K^{T}_{\mu
\rho}(k)\,K^{T\rho}_{\nu}(k)&=& - \,\frac {(k^{2}-M^{2})^{2}}
{(k^{2})^{2}}\, D_{\mu\nu}(k)\ . \end{eqnarray}

For completeness we note that
\begin{equation} \sum_{(\alpha)} \left[
E^{(\alpha) L}_{\mu}\,E^{(\alpha) L}_{\nu} + E^{(\alpha)
L}_{\mu}\,E^{(\alpha) T}_{\nu} + E^{(\alpha) T}_{\mu}\,E^{(\alpha)
L}_{\nu}\right]= K^{L}_{\mu\nu}(k) +\frac{M^{2}}{k^{2}}
\,\left(g_{\mu\nu}-\frac{k_{\mu}k_{\nu}}{k^{2}}\right)
\end{equation}
while
\begin{equation}
\sum_{(\alpha)} E^{(\alpha) T}_{\mu}(k)\,E^{(\alpha) T}_{\nu}(k) =
K^{T}_{\mu\nu}(k)- \frac{M^{2}}{k^{2}}
\,\left(g_{\mu\nu}-\frac{k_{\mu}k_{\nu}}{k^{2}}\right)\ .
\end{equation}

  \subsection{The Interaction Hamiltonian}

The interaction Hamiltonian, in LC gauge $A_{-}=0$, is derived to be
\begin{eqnarray}
&-&{\cal H}_{int}={\cal L}_{int} \nonumber \\ &=&
e\,M\, \,A_{\mu}A^{\mu}\,h - \frac{e\, m_{h}^{2}}{2\,M}\,(\eta^{2}+
h^{2})\, h + e \, (h \,\partial_{\mu}\eta-\eta \,\partial_{\mu} h)\,
A^{\mu} +\frac {e^{2}}{2} ( h^{2}+\eta^{2})\,A_{\mu}A^{\mu}\,\nonumber
\\ &-& \frac{\lambda}{4}\, (\eta^{2}+ h^{2})^{2}- \frac {e^{2}}{2}\,
\,\, \left(\frac{1}{\partial_{-} \,} j^{+}\right)\,
\left(\frac{1}{\partial_{-} \,}j^{+}\right)
\end{eqnarray}
where $j_{\mu}= (h \,\partial_{\mu}\eta-\eta
\partial_{\mu}\, h)$.  The last term here is the additional quartic
instantaneous interaction in the LF theory quantized in the LC gauge
(Appendix B).  No new instantaneous cubic interaction terms are
introduced.  The massive gauge field, when the mass is generated by the
Higgs mechanism, is described in our LC gauge framework by the
independent fields $A_{\perp}$ and $\eta$; the component $A_{+}$ is
dependent one.

  \section{ The GWS Model of Electroweak Interactions}

  \subsection{\boldmath The Quantization of the $SU(2)\otimes U(1)$
  Non-Abelian  Higgs Model in LC Gauge \unboldmath}

A condensed review of the GWS model will be given below to define our
notation.  The model constructs a unified description of the
electromagnetic and weak interactions by employing the spontaneously
broken gauge theory based on the nonabelian gauge group
$SU_{W}(2)\otimes U_{Y}(1)$, the direct product of the {\it weak }
isospin and the abelian {\it hypercharge} groups.  The corresponding
hermitian generators are $ ({\vec t}$ and $ t_{Y})$ respectively with
$\,{\vec t }=(t_{1},t_{2}, t_{3}) $, and $t_{Y}= Y \, I $.  Here ${\vec
t}$ are isospin generators, $I$ is the identity matrix, and $Y$
indicates the {\it hypercharge}.  For the spontaneous breaking a complex
scalar field, Higgs doublet $\Phi$, in the iso-spinor representation,
with $t=1/2, \,\,{\vec t } = {\vec \sigma}/2$, is introduced
\begin{equation}
\Phi= \left(\begin{array}{c} G^{+} \\
\chi^{o}\end{array} \right)\ .
\end{equation}
The value $Y(\Phi)=1/2$ is
assigned to it by convention such that the upper component $G^{+}$
corresponds to the unit eigenvalue of the ($U(1)_{em}$ or Charge)
generator $Q= (t_{3}+Y)$ and the lower one to the value zero.  Under
$SU_{W}(2)\otimes U_{Y}(1)$ it transforms as
\begin{equation} \Phi(x)
\to e^{ig\,{\vec t}\cdot {\vec \alpha(x)}}\, e^{ig'\,
t_{Y}\alpha_{Y}(x)}\,\, \Phi(x)
\end{equation}
where $g$ and $g'$
indicate the two gauge coupling constants while $\alpha_{a}(x)$ are the
gauge transformation parameters.  The gauge covariant derivative may be
defined as
\begin{equation} {\cal D}_{\mu} = (I\, \partial_{\mu}-ig
\,{\vec A}_{\mu}\cdot {\vec t} -ig' \,Y \,I\, B_{\mu})\,
\end{equation}
where ${\vec A}_{\mu}$ and $B_{\mu}$ are real valued gauge fields.

The nonabelian gauge theory Lagrangian is written as
\begin{equation}
{\cal L}= -\frac{1}{4}\, F^{a}_{\mu\nu}F_{a}^{\mu\nu} - \frac{1}{4}\,
F^{Y}_{\mu\nu}F_{Y}^{\mu\nu} + ({\cal D}_{\mu}\Phi)^{\dag}{{\cal
D}_{\mu}\Phi} -V(\Phi^{\dag}\Phi)
\end {equation}
where the gauge
invariant scalar potential contains, at most, quartic terms in $\Phi$,
so that the theory is renormalizable
\begin{equation} V(\Phi)= \mu^{2}
\, \Phi^{\dag}\Phi + \lambda (\Phi^{\dag}\Phi)^{2}
\end{equation}
where
$\lambda >0 $ and $ \mu^{2} < 0$.  The gauge field strengths are $\,
F^{a}_{\mu\nu}=\partial_{\mu} A^{a}_{\nu}-\partial_{\nu} A^{a}_{\mu}
+g\,f_{abc} A^{b}_{\mu}\, A^{c}_{\nu}\,$ where $a,b,c =1,2,3$ are the
$SU(2)$ gauge group indices, $f_{abc}\equiv \epsilon_{abc}$, while
$\,F^{Y}_{\mu\nu}=\partial_{\mu} B_{\nu}-\partial_{\nu} B_{\mu}$.

The description \cite{pre4} of SSB in the abelian case (Appendix A) can
be extended to the nonabelian one straightforwardly.  It may be shown
\cite{pre5} here too that none of the symmetry generators break the LF
vacuum symmetry, but the expression which counts the number of Goldstone
bosons is found to be identical to the one in the conventional theory
\cite{pesk}.  On the LF the tree level theory of the non-abelian Higgs
mechanism is straightforward to construct \cite{pre5}.  Its quantization
in the LC gauge parallels closely to that of the abelian Higgs theory.

It is convenient again to introduce real fields $\,h, \,\phi_{1}, \,
\phi_{2}, \,\phi_{3}\equiv G^{o}\,$ which have vanishing vacuum
expectation values and write
\begin{eqnarray}
G^{+}&\equiv& {-i\,
\phi^{-}}= -\frac{i}{\sqrt 2}\left(\,\phi_{1}(x)-i\,\phi_{2}(x)\right)
\nonumber \\ \chi^{o} &=& \frac{v}{\sqrt 2}+ \frac{1}{\sqrt 2} \left(
\,h(x)+ i\, G^{o}(x) \,\right)
\end{eqnarray}
where
$v=\sqrt{-\mu^{2}/\lambda}$.  In other words $\,\Phi=\Phi_{cl}
+\varphi\,$ such that
\begin{equation} \Phi_{cl}\equiv \VEV{0| \Phi |0} =
\frac{1}{\sqrt 2} \, \left(\begin{array}{c} 0 \\ v \end{array} \right)
\end{equation}
which is taken to be the classical vacuum configuration\footnote{
The stability of the asymmetric solution while the instability of
the symmetric one may be inferred from the study of the dynamical
(partial differential) equations of motion as usual.  } in the SSB
case when $\mu^{2}<0$.  This parameterization of $\Phi_{cl}$ can
always be assumed if we make use of the (global) symmetry of the
action under $SU_{W}(2)$ and $U_{Y}(1)$.  We verify that $\,t_{a}
\, \Phi_{cl} \neq 0$ but $ \, Q\Phi_{cl}\equiv (t_{3}+Y)=0$ where
the linear combination $Q$ is the generator of the unbroken
residual $U(1)_{em}$ symmetry.  We note also that $\Phi^{\dag}\Phi
= \left( \phi^{2}_{1}+ \phi^{2}_{2} +\phi^{2}_{3}+ \sigma^{2}
\right)/2$ where, $\sigma = (\,v + h(x)\,)$.  The potential $V$
defined above is invariant under the larger $O(4)\approx
SU(2)\times SU(2)$ symmetry, which is broken by the field $\sigma$
when it acquires a non zero vacuum expectation value.

The gauge field combinations $(W^{\pm}_{\mu}, \,Z)$ and photon $A_{\mu}$
(see below) are useful
\begin{eqnarray} W^{\pm}_{\mu} &=& \frac{1}{\sqrt
2}\, (A^{1}_{\mu}\, \mp \,i\, A^{2}_{\mu}) \nonumber \\ Z_{\mu}
&=& (A^{3}_{\mu} \, \cos \,\theta_{W}\, - B_{\mu}\, \sin
\,\theta_{W} ) \nonumber\\ A_{\mu} &=& (B_{\mu} \, \cos
\,\theta_{W}\,+ A^{3}_{\mu}\, \sin \,\theta_{W} )\ .
\end{eqnarray} Here $\,\theta_{W}$ is the Weinberg
angle such that $\,g \,\sin \,\theta_{W}= g' \, \cos \,\theta_{W} = e\,$
and $e$ is the electronic charge.  The gauge covariant derivative may be
conveniently re-expressed as
\begin{equation}
{\cal D}_{\mu}=
\partial_{\mu}-i\, \frac {g}{\sqrt 2}\, (\,W^{+}_{\mu}\, t_{+} +
W^{-}_{\mu}\, t_{-}\,) -i \frac {g}{\cos \,\theta_{W}} \, Z_{\mu}\, (\,
t_{3} - Q\, \sin^{2} \,\theta_{W}\,) -i\, e\,Q \, A_{\mu}
\end{equation}
where $Q=(\,t_{3}+Y\,)$ indicates the electric charge and $ t_{\pm}=
(t_{1}\pm i\,t_{2})= (\sigma_{1}\pm i\,\sigma_{2})/2 $. We find
\begin{equation}
{\cal D}_{\mu} \Phi= \left(\begin{array}{c}
\partial_{\mu}G^{+}-i m_{W}W^{+}_{\mu}-i [\frac{g
\,\cos(2\,\theta_{W})}{2\, \cos \theta_{W}}\, Z_{\mu} + e A_{\mu}]\,
G^{+}- \frac{ig}{2}\, W^{+}_{\mu} \,(h+iG^{o}) \\ \\ \frac{1}{\sqrt
2}\partial_{\mu}(h+iG^{o})+\frac{ig}{\sqrt 2}\, m_{Z}\,Z_{\mu} -
\frac{ig}{\sqrt 2}\, W^{-}_{\mu} G^{+} + \frac{ig}{\sqrt 2}\,\frac {1}{2
\cos \,\theta_{W}}Z_{\mu} \,(h+i G^{o}) \end{array} \right)
\end{equation}
while $({\cal D}^{\mu}\Phi)^{\dag}{{\cal D}_{\mu}\Phi}= $
\begin{eqnarray}
&& \vert\partial_{\mu}G^{+}-i m_{W}W^{+}_{\mu} -i
[\frac{g \,\cos(2\,\theta_{W})}{2\, \cos \,\theta_{W}}\, Z_{\mu} +
e A_{\mu}]\, G^{+}- \frac{ig}{2}\, W^{+}_{\mu} \,(h+iG^{o})
\vert^{2} \nonumber \\ &&+\,\frac{1}{2}\,\vert
\partial_{\mu}(h+iG^{o})+{ig}\, m_{Z}\,Z_{\mu} - {ig}\,
W^{-}_{\mu} G^{+} + {ig}\,\frac {1}{2 \cos \,\theta_{W}}Z_{\mu}
\,(h+i G^{o}) \vert^{2}\ .
\end{eqnarray}
Also
\begin{eqnarray}
V&=& \frac{1}{2}m^{2}_{h}\, h^{2}+ 2 \lambda v\, \left[
G^{+}G^{-} +\frac{1}{2}( G^{o 2} + h^{2})\right]\, h+
\lambda\left[G^{+}G^{-}+ \frac{1}{2}( G^{o 2}+
h^{2})\right]^{2}\nonumber \\ &=&\lambda \,\left[ G^{+}G^{-}
+\frac{1}{2}( G^{o 2} + h^{2}) +v\,h+
\frac{v^{2}}{2}+\frac{\mu^{2}}{2\lambda} \right]^{2}
\end{eqnarray}
where we set $\,m_{W}=gv/2, \, m_{Z}=m_{W}/ \cos\,\theta_{W}$
indicating the vector boson masses.  Interaction vertices are the
cubic and quartic terms in these expressions.  For example, the
cubic Higgs boson interaction with charged vector bosons is
\begin{equation}
\left [ g\,m_{W}\, W^{-}_{\mu} W^{+\,\mu} -i\,
\frac{g}{2} [(\partial_{\mu}G^{-})\, W^{+\,\mu}-
(\partial_{\mu}G^{+}) W^{- \,\mu}] +2 \lambda v\, G^{+}G^{-}
\right]\, h.
\end{equation}

The quadratic terms in the bosonic Lagrangian which define the free
theory are
\begin{eqnarray}
-\frac{1}{4}\,\left(\partial_{\mu}A_{\nu}-\partial_{\nu}A_{\mu}\right)^{2}&&
\nonumber \\
-\frac{1}{4}\,\left(\partial_{\mu}Z_{\nu}-\partial_{\nu}Z_{\mu}
\right)^{2} &+& \frac{1}{2}\,m^{2}_{Z}\, Z_{\mu}Z^{\mu}
+\frac{1}{2}(\partial^{\mu}G^{o})\,\partial_{\mu}G^{o}+ m_{Z}\,
Z_{\mu}\, \partial^{\mu} G^{o} \nonumber \\ -\frac{1}{2}\,(
\,\,\partial_{\mu}W^{+}_{\nu}-\partial_{\nu}W^{+}_{\mu} &)
(&\partial^{\mu}W^{-\nu}-\partial^{\nu}W^{-\mu}\,\,) + m^{2}_{W}
W^{-}_{\mu} W^{+\,\mu}\nonumber \\ &+&
(\partial_{\mu}G^{-})\,\partial^{\mu}G^{+} -i\, m_{W}
[(\partial_{\mu}G^{-})\, W^{+\,\mu}- (\partial_{\mu}G^{+})
W^{-}]\nonumber \\ &+&\frac {1}{2}
(\partial^{\mu}h)\,\partial_{\mu}h - \frac {1}{2} \, m^{2}_{h}\,
h^{2}\ .
\end{eqnarray}

No mass terms arise for the (Goldstone) fields $ G^{\pm}$ and
$G^{o}$ or for the photon field $A_{\mu}$.  We note the {\it tree
level} relations $(m_{h}/m_{W})^{2} = 8 \lambda/g^{2}$ and
$m^{2}_{h}/ m_{W}= (4/g)\,\lambda \, v\,$, $m^{2}_{W}/(m^{2}_{Z}
\, \cos ^{2} \,\theta_{W}) =1$, $(v/{\sqrt 2})= ({\sqrt
8}G_{F})^{-1/2} \approx 174$ GeV, and $\, G_{F}/\sqrt{2}=
g^{2}/(8 m^{2}_{W})=1/(2\,v^{2})$.  The bi-linear terms
corresponding to the charged fields may be rewritten in terms of
the real field components as\footnote{
\begin{eqnarray} \frac
{1}{\sqrt 2} \, (F^{1}_{\mu\nu} \mp i\, F^{2}_{\mu\nu})&=&
\partial_{\mu}W^{\pm}_{\nu}-\partial_{\nu}W^{\pm}_{\mu} \pm i\, g\,
\left(W^{\pm}_{\mu} A^{3}_{\nu}- W^{\pm}_{\nu}
A^{3}_{\mu}\right)\nonumber \\ F^{Y}_{\mu\nu}&=
&\left[(\partial_{\mu}A_{\nu}-\partial_{\nu}A_{\mu}) \, \cos \,
\theta_{W} -(\partial_{\mu}Z_{\nu}-\partial_{\nu}Z_{\mu}) \, \sin \,
\theta_{W} ) \right] \nonumber \\
\left(\partial_{\mu}A^{3}_{\nu}-\partial_{\nu}A^{3}_{\mu}\right)&=&
\left[(\partial_{\mu}Z_{\nu}-\partial_{\nu}Z_{\mu}) \, \cos \,
\theta_{W} +(\partial_{\mu}A_{\nu}-\partial_{\nu}A_{\mu}) \, \sin \,
\theta_{W} ) \right]
\end{eqnarray}}
\begin{eqnarray}
-\frac{1}{4}\,\left(\partial_{\mu}A^{1}_{\nu}-\partial_{\nu}A^{1}_{\mu}
\right)^{2} &+& \frac{1}{2}\,m^{2}_{W}\, A^{1}_{\mu}A^{1\mu}
+\frac{1}{2}(\partial^{\mu}\phi_{1})\,\partial_{\mu}\phi_{1}+ m_{W}\,
A^{1}_{\mu}\, \partial^{\mu} \phi_{1} \nonumber \\
-\frac{1}{4}\,\left(\partial_{\mu}A^{2}_{\nu}-\partial_{\nu}A^{2}_{\mu}
\right)^{2} &+& \frac{1}{2}\,m^{2}_{W}\, A^{2}_{\mu}A^{2\mu}
+\frac{1}{2}(\partial^{\mu}\phi_{2})\,\partial_{\mu}\phi_{2}+
m_{W}\, A^{2}_{\mu}\, \partial^{\mu} \phi_{2}\ .
\end{eqnarray}

The quantization in the LC gauge, $\,A_{-}=
Z_{-}=W^{\pm}_{-}\,=0$, is now straightforward.  We take over the
discussion in Section 2 on the abelian Higgs theory and the one
given in the earlier paper \cite{Srivastava:2000cf} on QCD for the
massless gauge field.  For comparison, we recall that the
conventional $R_{\xi}$ gauges in the equal-time framework requires
us to include in the theory also the ghost fields, which interact
with the Higgs and other physical fields.  Moreover, $W_{\mu}$,
$Z_{\mu}$, and $A_{\mu}$ may carry different parameters $\xi^{W}$,
$\xi^{Z}$, and $\xi^{\gamma}$ respectively in the gauge-fixing
terms.  The renormalization of these parameters also has to be
taken into consideration, and it is required to show that the
physical amplitudes do not depend on them.  The LC gauge framework
being discussed contains no ghost fields.  The 't Hooft conditions
corresponding to the massive vector bosons read as:
$\,\partial\cdot W^{\pm}= \pm i \, m_{W}\, G^{\pm}$, $ \,
\,\partial\cdot Z = m_{Z}\, G^{o},$ while for the massless field
we obtain \cite {Srivastava:2000cf} the Lorentz condition
$\partial\cdot A=0$.  The momentum space expansions of the
quantized field operators are easily found to be
\begin{eqnarray}
A^{\mu}(x)&=&{1\over {\sqrt {(2\pi)^{3}}}} \int d^{3}k
{\theta(k^{+})\over {\sqrt {2k^{+}}}}\, \sum_{(\perp)}
{E^{\mu}_{(\perp)}}(k) \left[a_{(\perp)}( k) e^{-i{ k}\cdot{x}} +a^{\dag
}_{(\perp)}(k) e^{i{ k}\cdot{ x}} \right ] \nonumber \\
W^{+}_{\mu}(x)&=&{1\over {\sqrt {(2\pi)^{3}}}} \int d^{3}k \,
{\theta(k^{+})\over {\sqrt {2k^{+}}}}\, \sum_{(\alpha)}
{E^{\mu}_{(\alpha)}}(k) \left[a^{W}_{(\alpha)}( k) e^{-i{ k}\cdot{x}} +
b^{W \dag }_{(\alpha)}(k) e^{i{ k}\cdot{ x}} \right ] \nonumber \\
Z^{\mu}(x)&=&{1\over {\sqrt {(2\pi)^{3}}}} \int d^{3}k \,
{\theta(k^{+})\over {\sqrt {2k^{+}}}}\, \sum_{(\alpha)}
{E^{\mu}_{(\alpha)}}(k) \left[a^{Z}_{(\alpha)}( k) e^{-i{ k}\cdot{x}}
+a^{Z\dag }_{(\alpha)}(k) e^{i{ k}\cdot{ x}} \right ]
\end{eqnarray}
where $d^{3}k \equiv d^{2}k^{\perp}dk^{+}$, $(\perp)= (1),(2)$, and
$(\alpha)= (\perp), (3)$.

For completeness, we collect here the cubic and quartic self
interactions of the gauge fields arising from the $ \, F^{a}_{\mu\nu}
F^{a \mu\nu}$ term
\begin{eqnarray}
&&ig\,
\left[(\partial_{\mu}W^{+}_{\nu}-\partial_{\nu}W^{+ }_{\mu})\, W^{- \mu}
- (\partial_{\mu}W^{-}_{\nu}-\partial_{\nu}W^{-}_{\mu})\,
W^{+\mu}\right]\, A^{3 \nu} \nonumber \\ && +{ig}\, \,
W^{+}_{\mu}W^{-}_{\nu}\,
\,(\partial^{\mu}A^{3\nu}-\partial^{\nu}A^{3\mu}) \nonumber \\ && +
g^{2}\left[\,\frac {1}{4}
(W^{+}_{\mu}W^{-}_{\nu}-W^{+}_{\nu}W^{-}_{\mu})^{2} -\,W^{+}_{\mu}\,
W^{-}_{\nu} \,A^{3}_{\rho}\,A^{3}_{\sigma}\, ( g^{\mu\nu} \,
g^{\rho\sigma} - g^{\rho\mu} \,g^{\sigma \nu})\right]
\end{eqnarray}
where $\,A^{3}_{\mu}= [ A_{\mu}\, \sin\, \theta_{W} + Z_{\mu}\,
\cos \, \theta_{W}]$.  Note that the complete $W^{+}W^{-}\gamma$
coupling, for example, includes the interaction terms carrying
$G^{\pm}$ fields arising from the $\vert {\cal D}_{\mu} \Phi
\vert^{2} $ term.

  \subsection{Fermionic Fields}

The LC gauge LF quantization when the fermionic fields are also present
is done by following closely the discussion \cite{Srivastava:2000cf,cov}
given in QCD.  The fermionic matter content of GWS model has three
generations with each one containing quarks and leptons.  The
left-handed components of the fermion fields are assigned to the
iso-spinor representation while the right-handed to the singlet of
$SU(2)_{W}$.  For example, in the first generation with quarks
$\,(u,\,d)\,$ and leptons $(\,\nu_{e},\, e^{-})$ we make the following
assignments
\begin{equation}
\psi_{L}:\qquad \left( \begin{array}{c}
\nu_{e} \\ e^{-} \end{array} \right)_{L}, \quad \left( \begin{array}{c}
u \\ d \end{array} \right)_{L}\quad \in \, t=\frac{1}{2}; \qquad \,
(u_{R},\, d_{R},\, e^{-}_{R})\quad \in \, t=0
\end{equation}
Here $\psi_{L}=[(1-\gamma_{5})/2]\psi $,
${\bar\psi}_{L}={\bar\psi}_{L}\, [(1+\gamma_{5})/2]$,
$\psi_{R}=[(1+\gamma_{5})/2]\psi $,
$\gamma_{5}=\gamma^{\dag}_{5}$, $\gamma^{2}_{5}=I$ etc.  Each
left-handed doublet is assigned a value of the hypercharge $Y$
similar to that of the Higgs doublet.  For example, $Y(u_{R})= Q
(u_{R})= Q(u_{L})= Q(u)=(Y+1/2)$ and $Q(d)=(Y-1/2)=Y(d_{R})$,
where $Y =Y(u_{L}) =Y(d_{L})$.  We recall $Y(e^{-}_{L})=-1/2$ and
$ Y(u_{L})=1/6$.

We base our discussion below on a single pair of generic fields
$\psi\equiv (u,\, d)^{T}$ with its left-handed components carrying the
hypercharge $Y$.  It may stand for $(\nu_{e}, e^{-})$, $(t,b)$, $
(c,s)$, etc.  The gauge invariant weak interaction Lagrangian for
massless fermions may be written as
\begin{equation} {\bar \psi}_{L}\, i
\, {\gamma^{\mu}\, {\cal D}_{\mu}}\, \psi_{L}+ {\bar u}_{R}\, i \,
{\gamma^{\mu}\, {\cal D}_{\mu}}\, u_{R}+ {\bar d}_{R}\, i \,
{\gamma^{\mu}\, {\cal D}_{\mu}}\, d_{R}\ .
\end{equation} The assignments
of the chiral components to distinct representations of $SU_{W}(2)$ and
the requirement of the gauge invariance do not allow one to introduce
directly the fermionic mass terms in the Lagrangian.  Such terms may,
however, be generated through SSB if the following gauge invariant
Yukawa interaction is added to the theory
\begin{equation}
-\lambda_{d}\, (\,{\bar \psi}_{L}\,\Phi \,)d_{R} -\lambda_{u} \,
(\,{\bar \psi}_{L}\,i \,\sigma_{2}\Phi^{*}\,) \,u_{R} + h.c.
\end{equation}
Here $\, \lambda_{u}, \, \lambda_{d}, $ are real
couplings, without any connection with the weak interaction coupling
constant, and we used $Y(\Phi^{*})=-1/2$.  We find the generation of the
mass terms:  $ \, -(m_{u}\, {\bar u} u+\, m_{d}\, {\bar d} d ) $, where
we set $\, \lambda_{d}\, v={\sqrt 2}\, m_{d}$, $\, \lambda_{u}\,
v={\sqrt 2} m_{u}$.  The Yukawa interaction terms are
\begin{eqnarray}
&& -\frac{g}{\sqrt 2} (\frac {m_{d}}{m_{W}})\,\left[\, {\bar
u}\frac{(1+\gamma_{5})}{2}\,d\, G^{+} +{\bar
d}\frac{(1-\gamma_{5})}{2}\, u\, G^{-} + \frac{1}{\sqrt 2}\, {\bar
d}d\, h + \frac {i}{\sqrt 2}\, {\bar d} \gamma_{5} d \, G^{o}
\right]  \\ && -\frac{g}{\sqrt 2} (\frac {m_{u}}{m_{W}})\,\left[\,
-{\bar u}\frac{(1-\gamma_{5})}{2}\,d\, G^{+} -{\bar
d}\frac{(1+\gamma_{5})}{2}\, u\, G^{-} + \frac{1}{\sqrt 2}\, {\bar
u}u\, h - \frac {i}{\sqrt 2}\, {\bar u} \gamma_{5} u \, G^{o}
\right]\nonumber
\end{eqnarray}
adding thereby additional parameters in the model.

The full fermionic Lagrangian is obtained from (55) and (56).  Besides
the Yukawa interactions in (57) it contains also the following terms
\begin{eqnarray}
&& {\bar u} \, \left[ i\, \gamma^{\mu}
(\partial_{\mu}-ieQ(u)A_{\mu})- m_{u}\right]\, u+ {\bar d} \, \left[i\,
\gamma^{\mu} (\partial_{\mu}-ie Q(d)A_{\mu})- m_{d}\right]\, d \nonumber
\\[1ex] && \qquad\qquad + g\,\left (\,W^{+}_{\mu}J^{\mu +}_{W}+
W^{-}_{\mu}J^{\mu -}_{W}+ Z_{\mu} J^{\mu }_{Z}\right)
\end{eqnarray}
where
\begin{eqnarray}
J^{\mu +}_{W}&=& \frac{1}{\sqrt 2} \left(\, {\bar
\psi}_{L} \gamma^{\mu}\, t_{+}\, \psi_{L}\, \right)=\frac{1}{2\sqrt 2}\,
{\bar u} \gamma^{\mu}\,(1-\gamma_{5}) d\, \nonumber \\ J^{\mu -}_{W}&=&
\frac{1}{\sqrt 2} \left(\, {\bar \psi}_{L} \gamma^{\mu}\, t_{-}\,
\psi_{L}\, \right)=\frac{1}{2\sqrt 2} \, {\bar d}
\gamma^{\mu}\,(1-\gamma_{5}) u\,\nonumber \\ J^{\mu}_{em}&=& Q(u)\,{\bar
u} \gamma^{\mu} \, u + Q(d)\,{\bar d} \gamma^{\mu}\, d \nonumber \\
J^{\mu}_{Z}&=& \frac{1}{\cos {\, \theta_{w}}} \left[\, {\bar
\psi}_{L}\,\gamma^{\mu}\, t_{3}\,\psi_{L}- \sin^{2}\, \theta_{W}\,
J^{\mu}_{em}\,\right] \nonumber \\ &=&\frac{1}{\cos {\, \theta_{w}}}
\left[\, \frac{1}{4}\,{\bar u}\,\gamma^{\mu}\, (1-\gamma_{5})\,u -
\frac{1}{4}\,{\bar d}\,\gamma^{\mu}\, (1-\gamma_{5})\,d - \sin^{2}\,
\theta_{W}\, J^{\mu}_{em}\,\right]
\end{eqnarray}
such that at the tree
level there are no flavor changing neutral currents.  The surviving
$U(1)_{em}$ gauge symmetry is also manifest.

The construction above gives the tree level description of the GWS
model in terms of the set of tree level parameters $(e, m_{W},
m_{Z}, m_{h}, m_{u}, m_{d})$ or alternatively $(e, \sin
\theta_{W}, v, m_{h}, m_{u}, m_{d})$.  The KM matrix can be
incorporated easily in our discussion.  The LF quantization of the
GWS model is performed following the discussions in Section 2,
Ref.  \cite{Srivastava:2000cf}, and the discussion in Appendix B.
The procedure closely follows the one adopted in connection with
the discussion \cite{Srivastava:2000cf} in LC gauge LF quantized
QCD.  In the GWS model we also have to take care in addition of
Yukawa interactions.  Besides the tree level interactions written
above, in the LF quantized theory we also have instantaneous
interaction in ${\cal H}^{LF}_{int}$ (see Appendix B).  They are
responsible for the restoration of the Lorentz covariance in the
computation of physical matrix elements etc.  The LF propagators
of the fields in LC gauge quantized GWS model are collected in
Appendix C.

  \section{ Illustrations}

  \subsection{\boldmath Decay $h\to W\,+\,W $ \unboldmath}

This decay is interesting also in connection with the Goldstone
boson or electroweak equivalence theorem.  It is clear from the
expressions of the relevant interaction vertices in Section 2 and
Section 3 that it suffices to consider the abelian theory.  The
$A\, A \, h$ interaction term gives the decay into two transverse
vector bosons.  The matrix element is
\begin{equation}
{\cal M}_{1}= (ieM) \, 2\, E^{(\alpha)}(k)\cdot E^{(\beta)}(k')=
-2ieM \,E^{(\alpha)}_{\perp}(k)\, E^{(\beta)}_{\perp}(k')\ .
\end{equation}
where $P_{\mu}= k_{\mu}+k'_{\mu}$
is the 4-momentum of the Higgs particle.  The $\eta^{2}\,h$ interaction
term produces longitudinal bosons in the Higgs decay.  The corresponding
matrix element is
\begin{eqnarray}
{\cal M}_{2}&=& -i\, \frac{\lambda v}{M^{2}}\, \, 2\, (i k\cdot
E^{(\alpha)}(k))\, (i k'\cdot E^{(\beta)}(k'))\, \nonumber \\
&=&i\, e \,\frac{m^{2}_{h}}{M}\,
\delta_{(\alpha)(3)}\delta_{(\beta)(3)}\ .
\end{eqnarray}
Finally, the
$\eta\, A\, h$ vertex gives
\begin{equation}
{\cal M}_{3}= - i\, \frac {e }{M} \, 2\, \left[ k^{\mu}\, k^{\nu}
+ k'^{\mu}\, k'^{\nu}+ k^{\mu}\, k'^{\nu}\right]\,
E^{(\alpha)}_{\mu}(k)\, E^{(\beta)}_{\nu}(k')\ .
\end{equation}
The total matrix element is
\begin{equation} {\cal
M}_{(\alpha)(\beta)}= 2\,i\,e\,M \left[ g_{\mu\nu} + \frac{1}{2}\,
\frac{m^{2}_{h}}{M^{4}}\, k_{\mu} \, k'_{\nu}- \frac{1}{M^{2}}\,
\left( k^{\mu}\, k^{\nu} + k'^{\mu}\, k'^{\nu}+ k^{\mu}\,
k'^{\nu}\right) \right] E^{(\alpha)}_{\mu}(k)\,
E^{(\beta)}_{\nu}(k')\ .
\end{equation}

Using mass-shell conditions we may rewrite
\begin{equation}
{\cal
M}_{(\alpha)(\beta)}= 2\,i\,e\,M \left[ g_{\mu\nu} + a\, k_{\mu} \,
k'_{\nu}+ b\, \left( k^{\mu}\, k^{\nu} + k'^{\mu}\, k'^{\nu} \right)\,
\right] E^{(\alpha)}_{\mu}(k)\, E^{(\beta)}_{\nu}(k')
\end{equation}
where $\, a= {(k \cdot k')}/{M^{4}}$ and $\, b= -1/M^{2}$.  It is
straightforward to compute the sum over polarizations of the squared
matrix element\footnote{ We use the simplifying properties of
$K_{\mu\nu}$, the relation $\, k_{\mu}\, k_{\nu}\, K^{\mu\nu}(k')=
-M^{2}+ 2\,(k\cdot k')\, k^{+}/k'^{+}$, and the mass-shell conditions.
} We find
\begin{equation} \sum_{(\alpha)}\sum_{(\beta)}\, \,\vert {\cal
M}_{(\alpha)(\beta)}\vert^{2} =(2\,e\,M)^{2}\, \left [ 2+
\frac{(k\cdot k')^{2}}{M^{4}}\right]
\end{equation}
which
agrees, as it should, with the result found when we use the
unitary (or Proca) gauge.

The discussion in the nonabelian theory of the Higgs decays into gauge
boson pair $W^{+}\, W^{-}$ is parallel to that of the abelian theory as
can be seen from the expressions in (35) and (48) of the corresponding
Higgs couplings.  We need only to replace $e\to g/2$ and $M\to m_{W}$ in
the discussion above.  We find
\begin{equation}
\sum_{(\alpha)}\sum_{(\beta)}\, \vert {\cal
M}_{(\alpha)(\beta)}\vert^{2} = \frac {g^{2}\,m^{4}_{h}}{4\,
m_{W}^{2}}\, \left [ 1 + 4\, \frac {m^{2}_{W}}{m^{4}_{h}}\,(
3\,m^{2}_{W}- m^{2}_{h})\right] \ .
\end{equation}
In the limit $m_{h} \gg m_{W}$ the leading term is the first one.
It derives solely from ${\cal M}_{2}$, {\em e.g.}, from the decay
to the would-be Goldstone particle $\eta$, as if we set the gauge
field as vanishing in the interaction Lagrangian.  Similar
discussions of other two body decays of the Higgs boson may be
given.

The additional contributions to the matrix element coming from the
would-be Goldstone bosons are found to be manifestly displayed.  The
matrix element ${\cal M}_{2}$, which derives solely from the would-be
Goldstone field, receives, compared to the others, an
$(m_{h}/m_{W})^{2}$ enhancement factor.  The result is general and has
been given the name of the Goldstone boson or electroweak equivalence
theorem \cite{cornwal}.  In the LF quantized theory it is revealed
transparently, and the physics of the longitudinal gauge bosons and
Higgs field can be described, under certain conditions, very well in
terms of the scalar self-interactions present in the initial Lagrangian
while ignoring the gauge fields.  This would not be true in the decay
under discussion if the mass of the Higgs boson is found, as currently
expected, to be around 115 GeV.  In fact, $[...]\approx [1+ 0.91]$ for
$m_{W}/m_{h}\approx 0.699$.

  \subsection{Muon Decay}

The cancellation of the noncovariant terms in the previous
illustration is seen easily also in muon decay, where the
noncovariant gauge propagator is involved.  However, in this case
we must also take into account a contribution from an
instantaneous interaction.

The terms in the interaction Lagrangian density responsible for the
process are read from (57), (58), and (59)
\begin{eqnarray}
&&\frac{g}{2\sqrt 2}\, [\,{\bar\nu}_{\mu^{-}}\, (1+\gamma_{5})\left(
\gamma \cdot W^{+} +\frac{i\,m_{\mu}}{m^{2}_{W}} \, \partial\cdot
W^{+}\right) \, \mu^{-} + \nonumber \\ && {\bar\mu}^{-} \, \left( \gamma
\cdot W^{-} -\frac{i m_{\mu}}{m^{2}_{W}} \, \partial\cdot W^{-}\right)
\,(1-\gamma_{5}) \, \nu_{\mu^{-}} + (\mu\to e)+ \cdots \,] \nonumber \\
&& + \mbox { quartic instantaneous interaction. }
\end{eqnarray}
Here we have made use of the 't Hooft conditions, $\,G^{\pm}=\mp i
(\partial\cdot W^{\pm})/m_{W} $ for convenience.  The matrix
element for the muon decay in momentum space, excluding the
instantaneous interaction contribution, reads as
\begin{eqnarray}
&&\left(\frac {ig}{2\sqrt 2}\right)^{2} \,\, {\bar u}(\nu_{\mu})\,
(1+\gamma_{5})\, \left(\gamma^{\mu}- \frac{m_{\mu}}{m^{2}_{W}}\,
k^{\mu}\right)\, u(\mu)\, \,\frac
{K_{\mu\nu}(k)}{(k^{2}-m^{2}_{W}+ i\epsilon)} \times\,\,\nonumber
\\ && {\bar u}(e)\, \, \left(\gamma^{\nu}-
\frac{m_{e}}{m^{2}_{W}}\, k^{\nu}\right)\,(1-\gamma_{5})
\,v({\bar\nu}_{e})
\end{eqnarray}
where $K_{\mu\nu}(k)$ is given
in (20).  On using simplifying properties of $K_{\mu\nu}(k)$
(Section 2) it reduces to (suppressing the constant factor)
\begin{eqnarray}
&&{\bar u}(\nu_{\mu})\, (1+\gamma_{5})\,
\gamma^{\mu} \, u(\mu)\, \frac {K_{\mu\nu}(k)}{(k^{2}-m^{2}_{W}+
i\epsilon)}\, {\bar u}(e)\,
\gamma^{\nu} (1-\gamma_{5})\, v({\bar\nu}_{e}) \nonumber \\
&-&\frac{m_{\mu}}{(k^{2}-m^{2}_{W}+ i\epsilon)\, k^{+}} \,{\bar
u}(\nu_{\mu})\, (1+\gamma_{5})\, \, u(\mu)\, \,{\bar u}(e)\, \gamma^{+}
(1-\gamma_{5})\, v({\bar\nu}_{e}) \nonumber \\
&-&\frac{m_{e}}{(k^{2}-m^{2}_{W}+ i\epsilon)\, k^{+}} \,{\bar
u}(\nu_{\mu})\, (1+\gamma_{5})\,\gamma^{+} \, u(\mu)\, \,{\bar
u}(e)\, (1-\gamma_{5})\, v({\bar\nu}_{e}) \nonumber \\ &+&
\frac{m_{\mu}m_{e}}{(k^{2}-m^{2}_{W}+ i\epsilon) \, m^{2}_{W}}
\,{\bar u}(\nu_{\mu})\, (1+\gamma_{5})\, \, u(\mu)\, \,{\bar
u}(e)\, (1-\gamma_{5})\, v({\bar\nu}_{e})\ .
\end{eqnarray}
Consider the contributions from the first term.  The noncovariant
terms carrying the $1/k^{+}$ dependence in $K_{\mu\nu}$ cancel the
second and the third terms.  Also an instantaneous contribution
comes from the last term in the expression of $K_{\mu\nu}$
\begin{equation}
- \frac{1}{k^{+ 2}}\, {\bar u}(\nu_{\mu})\, (1+\gamma_{5})\,
\gamma^{+} \, u(\mu)\, {\bar u}(e)\, \gamma^{+} (1-\gamma_{5})\,
v({\bar\nu}_{e})\ .
\end{equation}
It gets compensated by the
additional quartic instantaneous interaction term in our LC gauge
framework, which is easily derived by following the
straightforward procedure given in Appendix B.  The final result
agrees with the covariant matrix element found in the unitary
gauge.

  \subsection{\boldmath Decay $ t\to b\,+\, W^{+}$ \unboldmath}

The relevant interaction terms in the present case are
\begin{equation}
\frac{g}{2\sqrt 2}\,\, {\bar b}\,\left[ \, \gamma \cdot
W^{-}\,(1-\gamma_{5}) + \left( \frac{(m_{t}-m_{b})} {m_{W}} +
\frac{(m_{t}+m_{b})}{m_{W}}\, \gamma_{5}\right)\, G^{-}\right] \, t \,+
\,h.c.
\end{equation}
The matrix element may be written as
\begin{eqnarray}
&& \frac {ig}{2\sqrt 2}\, {\bar u}^{(r)}(b)\,\left[
\gamma^{\mu}\,(1-\gamma_{5})- \frac{m_{t}}{m^{2}_{W}}\, k^{\mu}\,
(1+\gamma_{5})\right]\, u^{(s)}(t)\, E^{(\alpha)}_{\mu} \nonumber
\\ &=& \frac {ig}{2\sqrt 2}\, {\bar u}^{(r)}(b)\,\left[ \gamma
\cdot E^{(\alpha)}(k)\,(1-\gamma_{5})+ \frac{m_{t}}{m_{W}}\,
\delta_{(\alpha)(3)}\, (1+\gamma_{5})\right]\, u^{(s)}(t)\ .
\end{eqnarray}
Here we have set $m_{b}=0$ for simplicity, and we
recall that $(\alpha)=(\perp), (3)$ indicate the three
polarization states of the massive vector boson as discussed in
Section 2.  For the spinor field we follow the notation of Ref.
\cite{Srivastava:2000cf}.  The $m_{t}$ enhancement of the matrix
element containing solely the would-be Goldstone bosons $G^{+}$ is
similar to that in the Higgs decay described above.  It is another
illustration of the electroweak equivalence theorem.  Since the
Higgs boson couples to fermion mass, the heavy fermion
contributions do not decouple.  The sum over spins and
polarizations of the squared invariant matrix element here is
found to be proportional to
\begin{eqnarray}
&&\left[ q^{\mu}p^{\nu}+ q^{\nu}p^{\mu} - (q\cdot p)
\,g^{\mu\nu} + (\frac{m_{t}}{m_{W}})^{2}\, \left( q \cdot p\,
\frac{k^{\mu}k^{\nu}}{m^{2}_{W}}\, -q^{\mu}k^{\nu}-
q^{\nu}k^{\mu}\right) \right]\, K_{\mu\nu}(k)\nonumber \\ &=&\left[
q^{\mu}p^{\nu}+ q^{\nu}p^{\mu} - (q\cdot p) \, g^{\mu\nu}\right] \,
d_{\mu\nu}(k) + (\frac{m_{t}}{m_{W}})^{2}\, {\left( q\cdot p- 2
m^{2}_{W} \,\frac{q^{+}}{k^{+}} \right)}
\end{eqnarray}
where the
mass-shell conditions such as $\, 2 k\cdot q=(m^{2}_{t}-m^{2}_{W})$, $\, 2
k\cdot p=(m^{2}_{t}+m^{2}_{W})$, $\,q^{2}=0$ have been used.  Collecting
together the noncovariant terms, we rewrite it as
\begin{eqnarray}
&=& -\left[ q^{\mu}p^{\nu}+ q^{\nu}p^{\mu} - (q\cdot p) \,
g^{\mu\nu}\right] \, g_{\mu\nu}+ \frac{1}{k^{+}}\, \left(
2\,q\cdot k\; p^{+}+ 2 k\cdot p\,q^{+} - 2 q\cdot p -2 m^{2}_{t}\,
q^{+}\right)\nonumber \\ && \quad +\,
(\frac{m_{t}}{m_{W}})^{2}\,q\cdot k \, = \,\, -\left[
q^{\mu}p^{\nu}+ q^{\nu}p^{\mu} - (q\cdot p) \, g^{\mu\nu}\right]
\, g_{\mu\nu}+\,(\frac{m_{t}}{m_{W}})^{2}\, q\cdot k \nonumber \\
&=&\left( -g_{\mu\nu} + \frac{k^{\mu}k^{\nu}}{m^{2}_{W}}\right) \,
\left[\, q^{\mu}p^{\nu}+ q^{\nu}p^{\mu} - (q\cdot p)\,
g^{\mu\nu}\right]\ .
\end{eqnarray}
The noncovariant terms cancel out giving the covariant
result of the unitary gauge\footnote{ $\Gamma= \frac{G_{F}\,
m^{3}_{t}}{8\sqrt{2\pi}}\, (1-\frac{m^{2}_{W}}{ m^{2}_{t}})^{2}\, (1+
2\, \frac{m^{2}_{W}}{m^{2}_{t}})$.}.

\section*{Conclusions}

The canonical quantization of LC gauge GWS electroweak theory in
the {\it front form} has been derived by using the Dirac procedure
to construct a self-consistent LF Hamiltonian theory. Combining
this with our previous work on light-front
QCD~\cite{Srivastava:2000cf}, we obtain a simultaneously unitary
and renormalizable gauge formulation of the Standard Model of the
strong and electroweak interactions.

The ghost-free interaction Hamiltonian of the Standard Model has
been obtained in a compact form by retaining the dependent
components $A_{+}$ and $\psi_{-}$ in the formulation.  Its form
closely resembles the interaction Hamiltonian of covariant theory,
except for the presence of a few additional instantaneous
interactions. Their derivation is given in  Appendix B. The
resulting Dyson-Wick perturbation theory expansion based on
equal-LF-time ordering is also constructed, allowing one to
perform higher-order computations in a straightforward fashion. In
contrast, in the conventional equal-time framework utilizing
 $R_{\xi}$ gauges, one is required to retain ghost fields which interact
with the physical fields.  Moreover, $W_{\mu}$, $Z_{\mu}$, and
$A_{\mu}$ can carry different parameters $\xi^{W}$, $\xi^{Z}$, and
$\xi^{\gamma},$ respectively, in the gauge-fixing terms.  The
renormalization of these parameters then also has to be taken into
consideration, and it is required to show that the physical
amplitudes do not depend on them. In view of the additional
simplifying properties of $K_{\mu\nu}$ and the (projector)
$D_{\mu\nu}$, computations in our framework require an effort
comparable to that of conventional covariant gauge theory.

In our LC gauge LF framework, the free massive gauge fields in the
electroweak theory satisfy simultaneously the 't Hooft conditions
as an operator equation. In the limit of vanishing mass of the
vector boson, the gauge field propagator goes over to the doubly
transverse gauge, ($n^{\mu}\, D_{\mu\nu}(k)=k^{\mu}\,
D_{\mu\nu}(k)=0$), the propagator found \cite{Srivastava:2000cf}
in QCD, in view of the Lorentz condition in the theory. As
discussed in Section 2, the factor $K_{\mu\nu}(k)$ in the gauge
propagator also carries important simplifying properties, similar
to the ones associated with the projector $D_{\mu\nu}(k)$.  The
transverse polarization vectors for massive or massless vector
boson may be taken to be $E^{\mu}_{(\perp)}(k)\equiv
-D^{\mu}_{\perp}(k),$ whereas the non-transverse third one in the
massive case is found to be parallel to the LC gauge direction $\,
E^{(3)}_{\mu}(k)= -(M/k^{+})\, n_{\mu}$.  Its projection along the
direction transverse to $k_{\mu}$ shares the spacelike vector
property carried by $E^{\mu}_{(\perp)}(k)$.

The Goldstone boson or electroweak equivalence
theorem~\cite{cornwal} becomes  transparent in our formulation.
Its content is illustrated in Section 4 by considering Higgs and
top decays. The computation of muon decay shows the relevance of
the instantaneous interactions for recovering manifest Lorentz
invariance in the physical gauge~\cite{schweda} theory framework.
They also correspond \cite{Srivastava:2000cf} to the
semi-classical (or nonrelativistic) limit frequently employed in
the conventional equal-time quantized theory.

The singularities in the noncovariant pieces of the field
propagators may be defined using the causal ML prescription for
$1/k^{+}$ when we employ dimensional regularization, as was shown
also in our earlier work on QCD. The power-counting rules in LC
gauge then become similar to those found in covariant gauge
theory.

We recall the explicit demonstration \cite{Srivastava:2000cf} of
the simplifying equality $Z_{1}=Z_{3}$ in QCD in our LC gauge
framework. Similar Ward identities are expected in the GWS model
as well.  These Ward identities simplify the task of computing
higher-loop corrections to physical processes.

Our light-front formulation of the Standard Model also provides
the basis for an ``event amplitude
generator"~\cite{Brodsky:2001ww} for high energy physics reactions
where each particle's final state is completely labelled in
momentum, helicity, and phase.  The application of the light-front
time evolution operator $P^-$ to an initial state will
systematically generate the tree and virtual loop graphs of the
$T$-matrix in light-front time-ordered perturbation theory.  In
our ghost-free light-cone gauge framework, the virtual loop
integrals only involve integration over the momenta of particles
with physical polarization and physical phase space $\prod
d^2k_{\perp i} d k^+_i$.  Renormalized amplitudes can be
explicitly constructed by subtracting from the divergent loops
amplitudes with nearly identical integrands corresponding to the
contribution of the relevant mass and coupling counter terms (the
``alternating denominator method")~\cite{Brodsky:1973kb}.


  \section*{Acknowledgments}

The hospitality offered to PPS at the Theory Division of SLAC and the
Theoretical Physics Department of the Fermilab is gratefully
acknowledged, as well as financial grants from the CAPES of Brazil
and the Proci\^encia program of UERJ, Rio de Janeiro.

  \bigskip

  \begin{center}{\Large \bf Appendix A}  \end{center}
  \bigskip

  \nl {\large\bf Spontaneous Symmetry Breaking Description on the LF}\par
  \bigskip


 We first consider, due to its relevance to
the discussion in Section 2,  the abelian case where
 the scalar theory Lagrangian with
$U(1)$ symmetry is given by
\begin{equation}
{\cal L}= {\partial}_{+}\phi^{\dag}\partial_{-}\phi+
{\partial}_{-}\phi^{\dag}\partial_{+}\phi -
\partial_{\perp}\phi^{\dag}\partial_{\perp}\phi -V(\phi^{\dag} \phi)
\end {equation}
where $V(\phi)= \mu^{2} \, \phi^{\dag}\phi + \lambda
(\phi^{\dag}\phi)^{2}$ with $\lambda >0 $ and $ \mu^{2} < 0$.  To
canonically quantize the theory we must construct an Hamiltonian
framework for the constrained dynamics described by the above
Lagrangian.  The Dirac procedure \cite{dir1} is convenient to use.
{\it Before} applying it, however, we make \cite{pre4} the
separation\footnote{ Such a decomposition may also be shown to
follow \cite{pre} as an external \cite{dir1} gauge-fixing
condition, corresponding to a first class constraint in the
theory, when we apply the Dirac procedure.  We note that $\int
d^{2} x^{\perp}dx^{-}\, \varphi=0$ such that $\varphi$ has
vanishing zero-longitudinal momentum-mode.}
\begin{displaymath}
\phi(\tau,x^{-},x^{\perp})=\omega(\tau,x^{\perp})+ \varphi(
\tau,x^{-},x^{\perp})
\end{displaymath}
 The field $\varphi $ indicates the
quantum fluctuations above the {\it dynamical
condensate} (or zero-longitudinal-momentum-mode) variable
$\omega(\tau, x^{\perp})$.
The LF Hamiltonian framework is found to contain in it also a
(second class) constraint equations \cite{pre4}, which relates the
condensate variables with the fluctuation fields.  The variable
$\omega$ is shown \cite{pre4,pre5, pre6} to have vanishing Dirac
brackets with itself and with $\varphi$.  It is thus a c-number
(background field) in the quantized theory\footnote{In the
 Schwinger model it is shown \cite{pre1c} to be a
q-number or an operator and where its presence gives rise to the
{\it chiral} and the $\theta$ or {\it condensate } vacua.  In the
case of the Chiral Schwinger model $\omega$ may be eliminated from
the theory by a field re-definition resulting in a different
degenerate vacuum structure.} The constraint
equations\footnote{They may \cite{pre4} also be obtained by
integrating the Lagrange equations but we must construct LF
Hamiltonian frame work to canonically quantize the theory.} in the
present case are \begin{equation} \int d^{2}x^{\perp}dx^{-}
\,[\partial_{\perp}\partial_{\perp} \phi- \frac{\delta V}{\delta
\phi^{\dag}}]=0, \qquad \int d^{2}x^{\perp}dx^{-}\,
[\partial_{\perp}\partial_{\perp} \phi^{\dag} - \frac{\delta
V}{\delta \phi}]=0.  \end{equation}
In the following discussion
we only consider the case where $\partial_{\perp}\omega =0$.  At the
classical (tree) level, since the fluctuations $\varphi$ are
assumed bounded, it follows \cite{pre4, pre5} that
$\,\delta V/\delta \phi \vert_{\phi=\omega}$ $=\delta V/\delta
\phi^{\dag} \vert_{\phi=\omega}=0$.  This coincides with the
result in the conventional equal-time framework.  It is obtained
there after imposing additional constraints, which are based on
physical considerations (seemingly not available or evident on the
LF).  The possible values of $\omega $ are $\omega=0$ or
$\omega^{\dag}\omega= -\mu^{2}/(2 \lambda)$.  The stability of
these solutions may be studied as usual from the Lagrange
equations; the nonvanishing $\omega $ gives rise to stable
solutions in the Nambu-Goldstone phase under study.  The
(classical) vacuum state is degenerate and characterized by a
fixed value of $\omega= \sqrt{-\mu^{2}/(2 \lambda)}\, e^{i\delta}
$ where $\delta$ is real and arbitrary.  In view of the invariance
of the action under the phase symmetry transformations:  $\,
\varphi\to e^{i\alpha} \varphi$, $\, \omega \to e^{i\alpha} \omega
$, we may, without any loss of generality, conveniently assume
$\omega\equiv v/{\sqrt 2}$ where $v= \sqrt{-\mu^{2}/\lambda}$ is a
fixed real constant.  A phase transformation would not leave
this classical vacuum state invariant, and the symmetry is said
to be broken spontaneously (see also Section 3.1)

At the quantum level, on the other hand, the LF field theoretic
generator of $U(1)$ symmetry annihilates the LF vacuum state, independent
of the broken symmetry or not.  The symmetry transformations always
leave the LF vacuum invariant, while the SSB is manifested, for example,
in the non-conservation of some of the symmetry currents
\cite{pre5,pre6}.
These features are true in general.

The Dirac procedure is straightforward to apply, and the quantized
theory is obtained by invoking the correspondence of the Dirac brackets
with the commutators of the corresponding quantized field operators.  In
the LF quantized theory we find the following non-vanishing
equal-$x^{+}$ commutator \begin{equation} [\varphi(x^{+}, x^{-},
x^{\perp}), \varphi(y^{+}, y^{-}, y^{\perp})]\,\vert_{x^{+}=y^{+}} =-
\frac{i}{4} \epsilon(x^{-}-y^{-})\, \delta^{2}(x^{\perp}-y^{\perp})
\end{equation} which does not violate the principle of microcausality on
the LF, in spite of the non-locality present in it along the $x^{-}$
direction.  The hermitian symmetry field theoretic generator is
constructed straightforwardly \begin{eqnarray} G(x^{+})&=& \int
d^{2}x^{\perp}dx^{-}\, j_{-}, \qquad\qquad \mbox{where} \nonumber \\
j_{\mu}&=& i\,\left[\varphi^{\dag}\, \partial_{\mu}\varphi- \varphi\,
\partial_{\mu} \varphi^{\dag}\right] \end{eqnarray} such that $\,
[\varphi (x), G]= \varphi$, $\, [\varphi (x)^{\dag}, G]= -
\varphi^{\dag}$.  The on-shell conserved Noether symmetry current is
given by \begin{equation} J_{\mu}= i\,\left[\phi^{\dag}\,
\partial_{\mu}\phi- \phi\, \partial_{\mu} \phi^{\dag}\right],
\qquad\quad \partial_{\mu}J^{\mu} = 0 \end{equation} which shows that
the symmetry current ($\phi=v/{\sqrt2} + \varphi$) \begin{eqnarray}
j_{\mu}&=&J_{\mu}- \frac {i\,v}{\sqrt 2} \, \partial_{\mu}(\varphi -
\varphi^{\dag})\nonumber \\ \partial^{\mu} j_{\mu} &=& \frac
{i\,v}{\sqrt 2} \, \partial\cdot\partial\,(\varphi-\varphi^{\dag})
\end{eqnarray} is not conserved in the broken phase.  In the LF
quantized theory, the two currents $j_{\mu}$ and $J_{\mu}$, however,
give rise to the same charge or
generator, if the surface terms may be ignored.

The LF commutator may be realized by the following momentum space
expansion \begin{equation} \varphi(x)= {1\over {\sqrt
{(2\pi)^{3}}}} \int d^{2}k^{\perp}dk^{+}\, {\theta(k^{+})\over
{\sqrt {2k^{+}}}}\, \left[\,a ( k) e^{-i{ k}\cdot{x}} +b^{\dag}(k)
e^{i{ k}\cdot{ x}} \right ] \end{equation} where the nonvanishing
commutators are $\, [a (k), a^{\dag}(l)]=$ $\, [b (k),
b^{\dag}(l)]$ $=\delta^{2}(k_{\perp}-l_{\perp})$
$\delta(k^{+}-l^{+})$.  The symmetry generator in momentum space
is found to be \begin{equation} G = \int d^{2}k^{\perp}dk^{+}\,
\theta(k^{+})\, \left[ a^{\dag}(k)a(k)- b^{\dag}(k)b(k)\right].
\end{equation} In the LF quantized theory only this term is
present.  It is already normal ordered and annihilates the LF
vacuum.  This is in contrast to the case of equal-time quantized
conventional theory, where there is an additional
term\footnote{In the equal-time quantized theory we have instead
$\partial_{t}(\varphi-\varphi^{\dag})$
in the expression of $j_{0}$ in (80).  It does not drop out
upon
coordinate space integration, and there is an additional term in the
corresponding generator which may not annihilate the vacuum state.  The
description of SSB \cite{pre6} is thus somewhat
different in the two {\it forms } of the theory.} in the field theoretic
symmetry generator which does not annihilate the corresponding
conventional vacuum state.  The LF vacuum thus remains invariant
under the symmetry transformations independent of the SSB in the
theory.  The broken symmetry manifests \cite{pre5} itself in the
non-conservation of (some) symmetry currents or in the operator LF
Hamiltonian.

\bigskip

\noindent{\large\bf Higgs mechanism in LF quantized theory}\cite{pre5}

\bigskip
The description below is relevant to the {\it front form} theory
of the GWS model in Section 3 which has a non-abelian Higgs
sector.

The SSB of continuous symmetry in the
non-abelian case is discussed in refs. \cite{pre5, pre6} by considering
an isospin-multiplet $\phi_{i}$,  $i=1,2,\cdots$,  of real scalar fields.
We {separate first} the {\it dynamical zero modes} or
{\it condensates} from the quantum fluctuations,
$\phi_{i}(\tau,x^{-},x^{\perp})=\omega_{i}(\tau,x^{\perp}) +
\varphi_{i}(\tau,x^{-},x^{\perp})$.  Then the Hamiltonian framework is
constructed
following the Dirac method.  We find in it, in addition to the commutators
and
the Hamiltonian, a set of coupled constraint equations.  At the tree level
they yield
$\,V^{'}_{i}(\omega)- \partial_{\perp}
\partial_{\perp}\, \omega_{i}=0 $.
For space independent $\omega$ we find the same expression as found in
the
conventional theory.

 It was also shown that the presence, in the
case of continuous SSB,  of the transverse directions was crucial for
showing
that the (dynamical) zero modes have vanishing Dirac brackets with the
non-zero
ones.  This furnishes us a new simple proof
 of the {\it Coleman theorem} on
the absence of Goldstone bosons in two dimensions, when we discuss
the SSB on the LF.

The field theoretic generators are now
$G_{a}= -i\, \int dx^{\perp}dx^{-}\, (\partial_{-}\varphi_{i})\,
(t_{a})_{ij}\,
\varphi_{j}$.  It is easily checked to be already normal ordered, as
in the
abelian case, and we need not impose it.  The symmetry generators on the
LF
thus annihilate the LF vacuum independent of the
form of the scalar potential and its symmetry is not broken.  We find
 $[\varphi_{i} (x), G_{a}]= (t_{a})_{ij}\,\varphi_{j}$,
 $[\omega_{i}, G_{a}]= 0$, and
 $[\, G_{a}, G_{b}\,]= if_{abc} \, G_{c}$ which is consistent with the
 generators annihilating the LF vacuum.  Not all the generators, however,
 commute with the Hamiltonian when SSB is present, say,
 when $\omega_{i}$
are determined from $\, (\lambda\, \omega_{i}\omega_{i}- m^{2})=0$.
 There may survive a residual unbroken symmetry if
 a set of linearly independent generators still commutes
 with the LF Hamiltonian.
 Such generators may be found by solving $({\tilde t}_{a})_{ij} \,
 \omega_{j}=0$ where ${\,\tilde t}_{a}$ are appropriate linearly
independent
 combinations, depending on the iso-vector $\omega=\{\omega_{i}\}$ chosen,
 of the matrix generators ${\,t}_{a}$ of the initial symmetry group.
 The corresponding generators $\,{\tilde G}_{a}$ commute with the
 Hamiltonian written in terms of $\varphi_{i}$ and fixed constants
 $\omega_{i}$.  The counting of the number of Goldstone bosons is thus
done
 as in the conventional theory.  The tree level Higgs Lagrangian is
 re-written  by the same procedure as
 in the conventional theory discussions, as done also in Section 3.
 The quantized theories of Higgs model though are different in the two
 {\it forms } of the theory as
 seen in Sections 2 and 3.

\vfill\eject

  \begin{center}{\Large \bf Appendix B}  \end{center}
  \bigskip

  \nl {\Large\bf
  Instantaneous Interactions in LF Quantized Theory }\par
  \bigskip

The additional instantaneous interactions in our LC gauge LF
theory framework in GWS model may be found straightforwardly by
following the procedure indicated in Ref.
\cite{Srivastava:2000cf}.  Such nonlocal terms are also required,
as shown there,
in order to restore the Lorentz covariance of
physical matrix elements.  They seem to have been missed in the
conventional theory discussions \cite{schweda,Srivastava:2000cf}
in noncovariant gauges.  It is worth stressing that they are
also present in {\it front form} Yukawa theory, which is not
even a gauge theory, as is shown below.  Some other illustrations
related to the abelian Higgs model, QCD, and the Yukawa couplings
in GWS model are also briefly described.  The instantaneous
interactions arise when we take into account the fact that the
nondynamical field components $\psi_{-}$ and $A_{+}$ are not independent
fields.  The {\it front form} theory framework, however,
permits us to re-express the interaction Hamiltonian in terms of
the full spinor and gauge fields, as previously shown in QCD.  It
results in an alternative ghost-free and practical framework, in view of
the Dyson-Wick expansion, for the computations in the
Standard model.  Unitarity and renormalizability
are also manifest.

  \bigskip
  \noindent {\large \bf LF quantized Yukawa theory}
  \bigskip

The LF quantization of the free spinor field was discussed in Ref.
\cite{cov} and the LF propagator of its dynamical component
derived; it was also shown not to contain any instantaneous term
in it.  We recall that in the {\it front form} theory the spinor
field\footnote{$x^{+} $ is taken as the LF-time while $(x^{-},
x^{\perp})$ indicate spatial coordinates.
 See, Refs.
\cite{cov,Srivastava:2000cf,pre} for notation and discussion on
the LF spinors.  We note: $\Lambda^{\pm}=\frac {1}{2}\,
\gamma^{\mp}\gamma^{\pm}$, $\gamma^{+}\psi_{-}=0$, etc.  } is
naturally decomposed into a dynamical field component
$\psi_{+}\equiv \Lambda^{+} \psi\,$ and a nondynamical auxiliary
field $\psi_{-}\equiv \Lambda^{-} \psi$, $\psi=
\psi_{+}+\psi_{-}$, where $\Lambda^{\pm}$, with $\Lambda^{+}+
\Lambda^{-}=1$, are hermitian projection operators.  Written in
the LF coordinates, the free Dirac Lagrangian may, in fact, be
re-written as
\begin{eqnarray}
{\cal L}^{o}&=& {\bar\psi}\, (i \gamma\cdot\partial
-m)\, \psi \nonumber \\ &=& {\bar\psi}\,(\Lambda^{+}
+\Lambda^{-})\,\frac{ \partial {\cal L}^{o}} {\partial\, {\bar\psi}}
\nonumber \\ &\to & \left.{\bar\psi}\,\Lambda^{-}\,\frac{ \partial {\cal
L}^{o}} {\partial\, {\bar\psi}}\, \right|_{\Lambda^{+}\frac{ \partial
{\cal L}^{o}} {\partial\, {\bar\psi}}=0} \nonumber \\
&=&{\bar\psi}_{+}\,(i \gamma\cdot\partial -m)\, \psi \quad\mbox{where}
\quad \gamma^{+}(i \gamma\cdot\partial -m)\, \psi =0 \nonumber \\ &=&
{\bar\psi}_{+}\, i\, \gamma^{+}\partial_{+}\, \psi_{+} +
{\bar\psi}_{+}\,(i\, \gamma^{\perp}\partial_{\perp} - m)\, \psi_{-}
\end{eqnarray}
Here we used $\, \Lambda^{\pm}\, \gamma\cdot\partial= (
\gamma^{\mp}\, \partial_{\mp}\, \Lambda^{\mp} +
\gamma^{\perp}\cdot\partial_{\perp}\, \Lambda^{\pm})\,$ which shows that
only $\psi_{+}$ is dynamical and independent field.  $\psi_{-}$ carries
no kinetic term and is a dependent field.  In fact, on taking the
variation of ${\cal L}^{o}$ with respect to the auxiliary field
${\bar\psi}_{-}$ we derive the constraint equation
\begin{equation}
{\Lambda^{+}\frac{ \partial {\cal L}^{o}} {\partial\, {\bar\psi}}=0},
\quad \mbox{\rm or}\quad \gamma^{+}\, (i\, \gamma\cdot \partial -
m)\,\psi=0
\end{equation}
which gives $\psi_{-}$
\begin{equation}
\psi_{-}=\frac{1}{2i\partial_{-}}\, (i\, \gamma^{\perp}\partial_{\perp}
+m) \, \gamma^{+}\, \psi_{+}
\end{equation}
showing it to be dependent
field component.

Consider now the {\it Yukawa theory} described by
\begin{eqnarray}
{\cal L}&=& {\bar\psi}\, (i \gamma\cdot\partial -m)\, \psi+
\frac{1}{2}\, (\partial_{\mu}\phi)^{2} -\frac{1}{2}\, M^{2}\phi^2+
g\, {\bar \psi}\psi\, \phi \nonumber \\ &=&
{\bar\psi}\,(\Lambda^{+} +\Lambda^{-})\,\frac{ \partial {\cal L}}
{\partial\, {\bar\psi}}+ \frac{1}{2}\, (\partial_{\mu}\phi)^{2}
-\frac{1}{2}\, M^{2}\phi^2\nonumber \\ &\to & \left.
{\bar\psi}_{+}\,\Lambda^{-}\,\frac{ \partial {\cal L}} {\partial\,
{\bar\psi}}\,\,\right|_{\Lambda^{+}\frac{ \partial {\cal L}}
{\partial\, {\bar\psi}}=0}+ \frac{1}{2}\, (\partial_{\mu}\phi)^{2}
-\frac{1}{2}\, M^{2}\phi^2 \ .
\end{eqnarray}
The nondynamical component $\,\psi_{-}\, $ is
now determined from the constraint equation
\begin{equation}
\Lambda^{+}\, \frac {\partial {\cal L}} {\partial\, {\bar\psi}} \equiv
\Lambda^{+}\,\, \left[\,(i\gamma\cdot\partial-m)\psi + S\,\right]=0
\quad \mbox{\rm or}\quad \gamma^{+}\, \,
\left[\,(i\gamma\cdot\partial-m)\psi + S\,\right]=0
\end{equation}
where
$S= g\, \phi \,\psi$.  We find
\begin{equation}
\psi_{-}\equiv
\Lambda^{-}\psi = \psi^{o}_{-}- \frac {1}{2i\partial_{-}} \,
\gamma^{+}\, S
\end{equation}
where we define
\begin{equation}
\psi^{o}_{-}= \frac{1}{2i\partial_{-}}\, \left(i\gamma^{\perp}\,
\partial_{\perp}+ m \right) \, \gamma^{+}\, \psi_{+} \ .
\end{equation}
Clearly,
\begin{equation}
\psi^{o}= \psi_{+}+ \psi^{o}_{-},
\end{equation}
where $\psi^{o}_{+}\equiv \psi_{+} =\Lambda^{+}\, \psi$,
satisfies the free field Dirac equation.  Also
\begin{equation}
\psi= \psi_{+} + \psi_{-}= \psi^{o}- \frac{1}{2i\, \partial_{-}}\,
\gamma^{+}\, S \ .
\end{equation}

The {\it front form} Yukawa theory Lagrangian reads as
\begin{eqnarray}
{\cal L} &=& {\bar\psi}_{+}\, \Lambda^{-}\,\,
\left[\,(i\gamma\cdot\partial-m)\psi + S\,\right] + \cdots \nonumber \\
&=& {\bar\psi}^{o}_{+}\,(i\gamma\cdot\partial-m)\psi^{o} -
{\bar\psi}^{o}_{+}\,(i\gamma\cdot\partial-m)\, \frac{1}{2i\,
\partial_{-}}\, \, \gamma^{+} \,S +{\bar\psi}^{o}_{+}\, S + \cdots
\nonumber \\ &=& {\cal L}^{o} + \, {\cal L}_{int}
\end{eqnarray}
where
\begin{eqnarray}
{\cal
L}^{o}&=&{\bar\psi}^{o}_{+}\,(i\gamma\cdot\partial-m)\psi^{o} +
\frac{1}{2}\, (\partial_{\mu}\phi)^{2} -\frac{1}{2}\, M^{2}\phi^{2}
\nonumber \\ {\cal L}_{int}&=& -
{\bar\psi}^{o}_{+}\,(i\gamma\cdot\partial-m)\, \frac{1}{2i\,
\partial_{-}}\, \, \gamma^{+} \,S +{\bar\psi}^{o}_{+} \, S \nonumber \\
&=& - {\bar\psi}^{o}\,(i\gamma^{\perp}\,\partial_{\perp}-m)\,
\frac{1}{2i\, \partial_{-}}\, \, \gamma^{+} \,S +{\bar\psi}^{o}
\,\Lambda^{-} \, S \nonumber \\ &\to & {\bar\psi}^{o} \,\Lambda^{+} \, S
+{\bar\psi}^{o} \,\Lambda^{-} \, S = \quad{\bar\psi}^{o} \, \, S
\nonumber \\ &=& g\, {\bar\psi}^{o} \,\left[ \psi^{o} - \frac{1}{2i\,
\partial_{-}}\, \gamma^{+}\, S \, \right] \, \phi \nonumber \\ &=& g\,
{\bar\psi}^{o} \, \psi^{o}\, \phi - g^{2}\, {\bar\psi}^{o} \,
\phi\, \frac{1}{2i\, \partial_{-}}\,\gamma^{+}\, \psi^{o}\, \phi \
.
\end{eqnarray}
In order to re-express the first term, we have
performed integrations by parts over the spatial coordinates
$x^{-},\, x^{\perp}$ in the Lagrangian; the $\gamma^{+}\,
\partial_{+}$ term drops out since $\gamma^{+\,2}=0$.  The
interaction, when expressed in terms of the free field $\psi^{o},$
contains an additional instantaneous term.  The LF quantization
may be performed straightforwardly and Dyson-Wick perturbation
theory expansion can be constructed.  It is worth recalling that the LF
fermionic propagator is also different from the one found in the
{\it instant form} quantized theory.  The instantaneous terms are
necessary, for example, in restoring the Lorentz invariance in the
computation of the meson nucleon scattering in the Yukawa theory.
Ignoring it would lead to disagreement in the calculations of the
nucleon self-energy in the LF and conventionally quantized
theories.  Their importance in LF quantized QCD in LC gauge was
also discussed in our earlier paper.

We remark that the expression of $\,\gamma^{+}\, S\,$ in Yukawa theory
contains only the dynamical $\psi_{+}$ component.
In the case of gauge theory, $\psi_{-} $ would occur also on the right
hand side of (88) if we
do not use the LC gauge, since $\, \gamma^{+}\, \gamma\cdot A,
\psi= 2\, A_{-}\, \psi_{-} +\gamma^{+}\, \gamma^{\perp}\, A_{\perp}\,
\psi_{+}$.

  \bigskip
  \noindent {\large \bf Abelian Higgs model}
  \bigskip

Next, we consider the derivation of the instantaneous interaction
terms in the abelian Higgs model discussed in Section 2.  From the
Lagrangian written in LF coordinates it is clear that $A_{+}$ is a
nondynamical since there is no corresponding kinetic term.  It is
also a dependent component.  Consider the equation of motion for
the gauge field
\begin{equation}
-\partial\cdot\partial \, A_{\mu} + \partial_{\mu} \,
(\partial\cdot A)= \frac {\partial {\cal L}}{\partial A^{\mu}} \ .
\end{equation}
We found significant simplifications in the fermionic
sector of LF quantized gauge theory if we adopt the LC gauge.
The
underlying gauge symmetry in the Higgs model allows one to adopt this
gauge, $A_{-}=0$.  From the expression of the Lagrangian (4) it then
follows that
\begin{equation} (
\partial\cdot A - M \,\eta)\,
\vert_{A_{-}=0} = e\, \frac {1}{\partial_{-}}\,K^{+} \quad \mbox{ where}
\quad K^{+}= \left.  \frac{1}{e}\,\frac {\partial {\cal L}}{\partial
A_{+}}\, \right|_{A_{-}=0, M=0}= (h\, \partial_{-}\eta -\eta \,
\partial_{-} h)
\end{equation}
Thus the free theory carries in it
simultaneously the 't Hooft condition, as was also demonstrated in the
Hamiltonian framework (and in the quantized theory).  When the SSB is
present and the mass of the gauge field is generated by the Higgs
mechanism in our framework, the massive gauge field is described by the
independent field components $A_{\perp}$ and $\eta$.
We may define,
as in the fermionic case, the dependent free field component
$A^{o}_{+}$ by the 't Hooft condition
\begin{equation}
\partial_{-}\,A^{o}_{+}= \partial_{\perp}\, A_{\perp}+ M\, \eta \
.
\end{equation}
It follows from (95) that
\begin{equation} A_{+}=
A^{o}_{+} + e\, \frac {1}{(\partial_{-})^{2}}\, K^{+}\ .
\end{equation}
Expressed in terms of the components $A_{\perp}, \eta, A^{o}_{+}$ and
$h$ the Lagrangian contains also instantaneous nonlocal interaction
terms.  They are indicated below on the right hand side of the arrow
corresponding to the term which gives rise to it
\begin{eqnarray}
M\, (A
\cdot \partial)\, \eta \,&\to & -\,e\, M\, \eta\, \frac
{1}{\partial_{-}}\, K^{+} \nonumber \\
e(h\partial_{\mu}\eta-\eta\partial_{\mu}h) A^{\mu}\, & \to & e^{2}\,
K^{+}\, \frac{1}{(\partial_{-})^{2}}\, K^{+} \nonumber \\ -\frac{1}{4}\,
F_{\mu\nu}\, F^{\mu\nu}\, &\to & e\, M\, \eta\, \frac
{1}{\partial_{-}}\, K^{+} -e^{2}\, \frac {1}{2}\, K^{+}\,
\frac{1}{(\partial_{-})^{2}}\, K^{+}
\end{eqnarray}
where integrations
by parts in the Lagrangian were freely used as in the fermionic case.
We observe that the cubic nonlocal interaction terms cancel leaving
behind only the quartic term.

    \bigskip
  \noindent {\large \bf LC gauge LF quantized QCD }
  \bigskip

In the fermionic piece we have now \begin{equation} S^{i}=
\gamma^{\mu}\, A^{a}_{\mu}\, ( t^{a})^{ij}\, \psi^{o j}\quad \mbox
{and} \quad \psi^{i}= \psi^{o i} - g\, \frac{1}{2i\,
\partial_{-}}\, \gamma^{+}\, S^{i}\, \vert_{A^{a}_{-}=0}
\ .
\end{equation}
For the nonabelian gauge field theory we follow
closely the above discussion for the Higgs model.  We have
\begin{equation} A^{a}_{+}= A^{o a}_{+} + g\, \frac
{1}{(\partial_{-})^{2}}\, j^{+ a}
\end{equation}
where in the
massless case we define $\,\partial_{-}A^{o a}_{+}
=\partial_{\perp} A_{\perp}$ and
\begin{eqnarray}
j^{+ a}&=
&\left.  \frac {1}{g}\, \frac {\partial {\cal L}}{\partial
A^{a}_{+}}\, \right|_{A^{a}_{-}=0} = f_{abc} \, A^{b}_{\perp}\,
\partial_{-}\, A^{c}_{\perp} + {\bar\psi}^{i}\,\gamma^{+}\,
(t^{a})^{ij}\, \psi^{ j}\nonumber \\ &=& f_{abc} \,
A^{b}_{\perp}\, \partial_{-}\, A^{c}_{\perp} + {\bar\psi}^{o
i}\,\gamma^{+}\, (t^{a})^{ij}\, \psi^{o j}\nonumber \\ &\equiv &
\left [\,K^{a} + L^{a}\, \right]\ .
\end{eqnarray}
The field
components $A^{a}_{+}$ and $\psi^{i}_{-}$ are again dependent
variables.  The fermionic piece contributes an instantaneous
seagull interaction as in the Yukawa theory.  There arises also
another type of instantaneous interaction
\begin{equation}
g^{2}\, L^{a}\, \frac{1}{ (\partial{-})^{2}}\, \left [\,K^{a} +
L^{a}\, \right]\ .
\end{equation}
Similar contribution coming from the gauge
field sector,
\begin{equation} -\frac {1}{4}\, F^{a\, \mu\nu}\, F_{a \, \mu\nu}= \frac
{1}{2}\, \left[ F_{a\,+ -}\, F_{a \, + -}+ 2\, F_{a\, + \perp}\, F_{a \,
- \perp}- \frac{1} {2}\, F_{a\, \perp\perp'}\, F_{a \, \perp\perp'}
\right]
\end{equation}
is found to be
\begin{equation} g^{2}\, K^{a}\,
\frac{1}{ (\partial{-})^{2}}\, \left [\,K^{a} + L^{a}\, \right]
-\frac {1}{2}\, g^{2}\, \left [\,K^{a} + L^{a}\, \right] \,
\frac{1}{ (\partial{-})^{2}}\, \left [\,K^{a} + L^{a}\, \right]\ .
\end{equation}

The interaction Hamiltonian in QCD follows:
\cite{Srivastava:2000cf}
\begin{eqnarray}
{\cal H}_{int}= -{\cal
L}_{int}&=& -g \,{{\bar\psi}}^{i} \gamma^{\mu}\, (t^{a})^{ij}\,
{{\psi}}^{j} \,A_{\mu}^{a} \nonumber \\ && +\frac{g}{2}\, f^{abc}
\,(\partial_{\mu}{A^{a}}_{\nu}- \partial_{\nu}{A^{a}}_{\mu}) A^{b\mu}
A^{c\nu} \nonumber \\ && +\frac {g^2}{4}\, f^{abc}f^{ade} {A_{b\mu}}
{A^{d\mu}} A_{c\nu} A^{e\nu} \nonumber \\ && - \frac{g^{2}}{ 2}\,\,
{{\bar\psi}}^{i} \gamma^{+}
\,\gamma^{\mu}\,A_{\mu}^{a}\,(t^{a})^{ij}\,\frac{1}{i\partial_{-}} \,
\gamma^{\nu} \,A_{\nu}^{b}\,(t^{b})^{jk}\,{\psi}^{k} \nonumber \\ &&
-\frac{g^{2}}{ 2}\, (\frac{1}{i\partial_{-}}\,j^{+}_{a})\,
(\frac{1}{i\partial_{-}}\,j^{+}_{a} )
\end{eqnarray}
where
\begin{equation}
j^{+}_{a}={{\bar\psi}}^{i} \gamma^{+} (
{t_{a}})^{ij}{{\psi}}^{j} + f_{abc} (\partial_{-} A_{b\mu}) A^{c\mu}
\end{equation}
and a sum over distinct quark and lepton flavors, not
written explicitly, is understood in (105) and (106).

   \bigskip
  \noindent {\large \bf  GWS model }
 \bigskip

In the electro-weak sector of the Standard model, $``\,S\,"$ contains
terms such as $\, \gamma^{\mu}\, Z_{\mu}\, \psi\, $ etc.  Only in the LC
gauge, with $A_{-}=Z_{-}= W^{\pm}_{-}=0$, the $\gamma^{+}\,S$ will
contain solely the dynamical $" + " $ component of the fermionic fields
involved.  The discussion in the GWS model in LC gauge follows closely
the one given in QCD.

\vfill\eject

  \bigskip

  \begin{center}{\Large \bf Appendix C}  \end{center}
  \bigskip

  \nl {\large\bf Feynman Rules and
   Propagators }\par
  \bigskip

The Dyson-Wick perturbation theory expansion on the LF can be
realized in momentum space by employing the Fourier transform of
the fields and the propagators discussed in Sections 2, 3, and in
Ref. \cite{Srivastava:2000cf}.  Many of the rules of the Feynman
diagrams, for example, the symmetry factor $1/2$ for gluon loop, a
minus sign associated with fermionic loops etc., are the same as
those found in the conventional covariant framework.
There are
some differences:$\,\,$ for example, the external quark line now
carries a factor $\theta(p^{+})\sqrt {{m}/{p^{+}}};$ the
external boson line carries the factor $\theta(q^{+})/\sqrt{2q^{+}}\;$
and
the Lorentz invariant phase space factor is $\int
d^{2}p^{\perp}dp^{+}\,\theta(p^{+})/(2p^{+})$.  The external
massive vector boson line carries the polarization vector $E^{\mu
(\alpha)}(q)$.  Its properties and the sum over the polarization
states are given Section 2.  The notation for the quark field is as
given in Refs. \cite{cov, Srivastava:2000cf}.  The instantaneous
interactions in electroweak theory may be found using Appendix B.
The momentum space vertices can be derived straightforwardly
employing the Fourier transforms of the fields given in the text
and illustrated in Ref. \cite{Srivastava:2000cf} in QCD.  The free
propagators are

\bigskip


  \nl {\it Fermionic  propagator}:
\[
i\,\delta_{ij} \, \frac{N(p)}{p^2-m^2+i\epsilon},
\quad {\rm with} \quad N(p)=({\not {p}}+m)- (p^{2}-m^{2})\,
\frac {\gamma^{+}}{2 p^{+}},\qquad \quad\epsilon> 0, \]
where $p_\mu$ is the quark 4-momentum and $i$ and $j$ are color indices.
The noncovariant second term on the right hand side is present
only in the propagator of the dependent field $\psi_{-}$.  Also
$\, N(p)= ({\not {p}}_{on}+m)\, $ where $p_{on}: \, \left(\,
(m^{2}+p^{2}_{\perp})/2p^{+}, p^{+}, p^{\perp}\, \right)$.


\bigskip \nl {\it Photon propagator}:
\[ i\, \frac {D_{\mu\nu}(q)}{q^{2}+ i\epsilon}, \quad {\rm with}
\quad \, D_{\mu\nu}(q)= \left( -g_{\mu\nu} + \frac{ n_\mu q_\nu +
q_\mu n_\nu }{ n\cdot q } -\frac{ q^2 }{ (n\cdot q)^2} n_\mu n_\nu
\right), \] where $q_\mu$ is the photon 4-momentum and $n_\mu$ is
the gauge direction.  We choose $n_\mu\equiv \delta^{+}_\mu$ and
$n^{*}_\mu\equiv \delta^{-}_\mu$, the dual of $n_\mu$.

  \bigskip
  \nl {\it Vector boson propagators}:

\begin{displaymath}
\VEV{ W^{+}_{\mu} (q)\,W^{-}_{\nu} (-q)} = i \frac
{K_{\mu\nu}(q)}{q^{2}-m^{2}_{W}+ i\epsilon}, \end{displaymath} where
\begin{displaymath} K_{\mu\nu}(q)= \left( -g_{\mu\nu} + \frac{ n_\mu
q_\nu + q_\mu n_\nu }{ n\cdot q } -\frac{ (q^2-m^{2}_{W}) }{ (n\cdot
q)^2} n_\mu n_\nu \right),
\end{displaymath} where $q_\mu$ is the vector
boson 4-momentum and $n_\mu$ is the gauge direction.  We choose
$n_\mu\equiv \delta^{+}_\mu$ and $n^{*}_\mu\equiv \delta^{-}_\mu$, the
dual of $n_\mu$.  For the neutral $Z$ vector boson $m_{W}$ is
substituted by $m_{Z}$.

The scalar fields $G^{\pm}, G^{o}$ and $ h $ have the standard covariant
propagators $ i / (q^{2}- M^{2})$ where $M= m_{W}, m_{Z}$ and $m_{h}$
respectively.

It is worth recalling \cite{Srivastava:2000cf} the procedure for
computing the discontinuity or imaginary parts of any
Feynman diagram,  employing the
Cutkosky rules in our LF framework.  For each cut,  replace
$\, 1/(p^{2}-m^{2}+i\,\epsilon)\, \to \, -2\pi\, i\,  \delta(p^{2}-m^{2})$
and then perform the loop integrals.  We note that
$(p^2-m^2)\,\delta(p^{2}-m^{2})=0$ such that last term in each of
$N(p)$, $D_{\mu\nu}(q)$, and $K_{\mu\nu}(q)$ gives vanishing contribution.

\end {document}